\documentstyle[aps]{revtex}

\newcommand{\disregard}[1]{}

\newcommand{\Dh}{D$_{\text{2h}}$}
\newcommand{\Dd}{D$_{\text{2}}$}
\newcommand{\DT}{D$_{\text{2h}}^{\text{T}}$}
\newcommand{\DTD}{D$_{\text{2h}}^{\text{TD}}$}
\newcommand{\DD}{D$_{\text{2h}}^{\text{D}}$}

\newcommand{\hatI}{\hat{I}}
\newcommand{\hatL}{\hat{L}}

\newcommand{\hatO}{\hat{O}}

\newcommand{\hatT}{\hat{T}}

\newcommand{\hatE}{\hat{E}}
\newcommand{\barE}{\bar{E}}
\newcommand{\hatK}{\hat{K}}
\newcommand{\hatP}{\hat{P}}

\newcommand{\hatU}{\hat{U}}

\newcommand{\hatUT}{\hat{U}^T}
\newcommand{\hatPT}{\hat{P}^T}

\newcommand{\hatQ}[1]{\hat{Q}_{#1}}
\newcommand{\hatM}[1]{\hat{M}_{#1}}
\newcommand{\hatR}[1]{\hat{R}_{#1}}

\newcommand{\hatS}[1]{\hat{S}_{#1}}

\newcommand{\hatRT}[1]{\hat{R}^T_{#1}}

\newcommand{\hatST}[1]{\hat{S}^T_{#1}}

\newcommand{\hatME}{\hat{\mathcal{E}}}
\newcommand{\barME}{\bar{\mathcal{E}}}
\newcommand{\hatMT}{\hat{\mathcal{T}}}
\newcommand{\barMT}{\bar{\mathcal{T}}}
\newcommand{\hatMP}{\hat{\mathcal{P}}}
\newcommand{\barMP}{\bar{\mathcal{P}}}
\newcommand{\hatMA}{\hat{\mathcal{A}}}

\newcommand{\hatMK}{\hat{\mathcal{K}}}

\newcommand{\hatMU}{\hat{\mathcal{U}}}
\newcommand{\barMU}{\bar{\mathcal{U}}}
\newcommand{\hatMPT}{\hat{\mathcal{P}}^T}
\newcommand{\barMPT}{\bar{\mathcal{P}}^T}
\newcommand{\hatMRT}[1]{\hat{\mathcal{R}}_{#1}^T}
\newcommand{\barMRT}[1]{\bar{\mathcal{R}}_{#1}^T}
\newcommand{\hatMST}[1]{\hat{\mathcal{S}}_{#1}^T}
\newcommand{\barMST}[1]{\bar{\mathcal{S}}_{#1}^T}
\newcommand{\hatMR}[1]{\hat{\mathcal{R}}_{#1}}
\newcommand{\barMR}[1]{\bar{\mathcal{R}}_{#1}}
\newcommand{\hatMS}[1]{\hat{\mathcal{S}}_{#1}}
\newcommand{\barMS}[1]{\bar{\mathcal{S}}_{#1}}

\newcommand{\hatBE}{\hat{\mathbf{E}}}
\newcommand{\hatBT}{\hat{\mathbf{T}}}
\newcommand{\hatBP}{\hat{\mathbf{P}}}

\newcommand{\hatBU}{\hat{\mathbf{U}}}
\newcommand{\hatBPT}{\hat{\mathbf{P}}^T}
\newcommand{\hatBRT}[1]{\hat{\mathbf{R}}_{#1}^T}
\newcommand{\hatBST}[1]{\hat{\mathbf{S}}_{#1}^T}
\newcommand{\hatBR}[1]{\hat{\mathbf{R}}_{#1}}
\newcommand{\hatBS}[1]{\hat{\mathbf{S}}_{#1}}

\newcommand{\hatsi}{\hat{\sigma}}

\newcommand{\hoket}[1]{|n_xn_yn_z,#1\rangle}

\newcommand{\demi}{\frac{1}{2}}
\newcommand{\tdemi}{{\textstyle{\demi}}}

\newcommand{\ps}{pseudo}

\newcommand{\ay}{anti}
\newcommand{\inv}{invariant}
\newcommand{\cov}[1]{{#1}-covariant}
\newcommand{\pscov}[1]{{#1}-{\ps}covariant}
\newcommand{\acov}[1]{{#1}-{\ay}covariant}
\newcommand{\apscov}[1]{{#1}-{\ay}{\ps}covariant}

\newcommand{\be}{\begin{equation}}
\newcommand{\ee}{\end{equation}}
\newcommand{\ba}{\begin{array}}
\newcommand{\ea}{\end{array}}
\newcommand{\bn}{\begin{eqnarray}}
\newcommand{\en}{\end{eqnarray}}
\newcommand{\bnl}{\begin{mathletters}\begin{eqnarray}}
\newcommand{\enl}{\end{eqnarray}\end{mathletters}}
\newcommand{\bml}{\begin{mathletters}}
\newcommand{\eml}{\end{mathletters}}
\newcommand{\bc}{\begin{center}}
\newcommand{\ec}{\end{center}}
\newcommand{\bi}{\begin{itemize}}
\newcommand{\ei}{\end{itemize}}

\newcommand{\bnll}[1]{\begin{mathletters}\label{#1}\begin{eqnarray}}
\newcommand{\enll}{\end{eqnarray}\end{mathletters}}

\tighten
\begin{document}

\draft
\twocolumn[\columnwidth\textwidth\csname@twocolumnfalse\endcsname
\title{Point symmetries in the Hartree-Fock approach:
       Densities, shapes and currents}

\author{
 J. Dobaczewski,$^{1,2}$
 J. Dudek,$^{2}$
 S.G. Rohozi\'nski,$^{1}$ and
 T.R. Werner$^{1,2}$
}

\address{
$^1$Institute of Theoretical Physics, Warsaw University,
    Ho\.za 69,  PL-00681, Warsaw, Poland                            \\
$^2$Institute de Recherches Subatomiques,
    CNRS-IN$_2$P$_3$/Universit\'e
    Louis Pasteur, F-67037 Strasbourg Cedex 2, France               \\
}

\maketitle

\begin{abstract}
Three mutually perpendicular symmetry axes of the second order,
inversion, and time reversal can be used to construct a double point
group denoted by {\DTD}. Properties of this group are analyzed in
relation to the symmetry and symmetry-breaking effects within the
mean-field (Hartree-Fock) theories, both in even and odd fermion
systems. We enumerate space symmetries of local one-body densities,
and symmetries of electromagnetic moments, that appear when some or all of
the {\DTD} elements represent self-consistent mean-field symmetries.
\end{abstract}

\pacs{PACS numbers: 21.60.-n, 21.60.Jz, 21.10.Ky}
\addvspace{5mm}]

\narrowtext

\section{Introduction}
\label{sec1}

Point-group symmetries play a very important role in nuclear
mean-field theories.  Two distinct aspects of this role can be
singled out. Firstly, the point symmetries of a Hamiltonian provide
good quantum numbers that can conveniently be used to label its
eigenstates. They help to formulate selection rules for
electromagnetic and/or other types of transitions, and allow for
solving the stationary problems in subspaces rather than in the
complete Hilbert space of the problem in question. In that respect
the use of point symmetries in nuclear physics resembles their use in
other branches of physics. Secondly, however, and this aspect is more
specific to nuclear structure domain, the use of the self-consistent
Hartree-Fock (HF) or Hartree-Fock-Bogolyubov (HFB) mean field
approximations invariably leads to the problem of self-consistent
symmetries and related spontaneous symmetry-breaking
mechanisms\cite{[RS80]}.

In this article we aim at describing properties of nuclear one-body
densities under the action of point symmetries. For the time-even
densities we calculate the electric multipole moments which
give information about nuclear shapes. Various point symmetries obeyed by
the hamiltonian lead then to various types of allowed shapes.
In addition,
for the time-odd densities we calculate magnetic multipole
moments which give information about current distributions in
nuclei, i.e., about the {"}shapes" of matter flow. Again, various
conserved symmetries restrict these flow patterns in different ways
that are studied in this paper.

Numerous experiments indicate that a great number of nuclei are
deformed in their ground states. Interpretation of the corresponding
results shows that most often the shapes involved are
axially-symmetric. Many realistic calculations, e.g., those based on
the nuclear mean-field approximation, reproduce these experimental
data. However, the same calculations suggest that the excited nuclear
states often correspond to the nucleonic mass distributions that have
the so-called triaxial shapes. It thus becomes clear that in a realistic
description of the nuclear properties, the spontaneous symmetry
breaking leading to the triaxially symmetric objects must be given
attention.

The classical point group that contains three mutually perpendicular
symmetry axes of the second order passing through a common point
is denoted {\Dd} (cf., e.g.,
Refs.~\cite{[Ham62],[Kos63],[Lan81],[Cor84]}). Attaching to {\Dd} the
three mutually perpendicular symmetry planes spanned on the symmetry axes
gives the {\Dh} point group
which contains all the spatial symmetries of interest in the present paper.

As it is well known, the classical ({\em single}) point groups can be
applied to spinless particles and/or systems of an even number of
fermions, and thus to even-even nuclei. However, for odd fermion
systems and, in particular, in the single-nucleon space, these have
no faithful irreducible representations. There exist two methods to
remedy this problem. One is to extend the notion of the group
representation and to introduce projective or ray representations
\cite{[Ham62],[Cur62],[Bra72]}. Another one, which is employed in
the present work, is to enlarge the single groups by adjoining the
rotation through angle $2\pi$ and all its products with the original
group elements, and to double in this way the order of the
group\cite{[Cor84]}.

Physically, the need of such an extension is related to the fact that
in the space of spinors the rotation through angle of $2\pi$
necessarily changes the sign of the wave
function of an odd-fermion system.
Since within the group theory a multiplication of
a group element by a number is not defined, the change of sign
must be introduced as an extra group element. The point group
enlarged in this way is called the {\em double} point group and usually
denoted with the superscript ``D'' (cf., e.g.,
Refs.~\cite{[Lan81],[Cor84]}), although some authors, see, e.g.,
\cite{[Kos63]}, denote single and double point groups by the same
symbols. Here we follow the former convention, and thus the double
group corresponding to single group {\Dh} is denoted by {\DD}.

In the case of classical objects the elements of a symmetry point
group are real orthogonal coordinate transformations. In quantum
mechanics it is often of advantage to take into consideration both
the spatial symmetries and the time-reversal operator explicitly
and treat them as elements of a common ensemble of symmetry operators.
Time-reversal symmetry operator (antilinear) and the space
symmetries (linear) have usually been considered separately
(cf., e.g., Ref.~\cite{[Kos63]}). Here we follow Ref.\cite{[Wig59]}
and add the time-reversal
operator to the set of group elements, thus obtaining new groups denoted
${\text{D}}_{\text{2h}}^{\text{T}}$ and
${\text{D}}_{\text{2h}}^{\text{DT}}$. Hence, the {\Dh} group
is a subgroup of {\DT}
composed of its linear elements, and similarly, the
{\DD} group is composed of the linear elements of {\DTD}.

Gauge symmetries, which pertain to pairing correlations of nucleons,
can be added independently and are not considered in the present
study. In particular, neutron-proton correlations are not discussed.
Therefore, the isospin degree of freedom is irrelevant in the
discussion and can be disregarded to simplify the notation.

In this paper our goal is threefold:  First, in Sec.~\ref{sec2}, we
present and discuss properties of the single group {\DT} and double
group {\DTD} that are appropriate for a description of even and odd
fermion systems, respectively. In particular, we recall the
classification of representations of both groups, and classify
properties of the group elements when they are represented in fermion
Fock space. Second, in Sec.~\ref{sec6}, we present explicit symmetry
properties of local densities with respect to the operators in the
{\DTD} group.  This problem has been solved in particular
applications \cite{[Bon87],[Dob97]}; however, it can be solved in
many different ways, and it is useful to have a systematic approach
which enumerates all available options. Although the local densities
are most important for applications using the local density
approximation (LDA), or those using the Skyrme effective interaction
(see respectively Refs.\cite{[Neg75],[Dre90]} or Ref.\cite{[Que78]}
for reviews), they also define general properties of average values
of any local one-body operator. Finally, in Sec.~\ref{sec8}, we
discuss symmetries of multipole moments which define the nuclear
shapes and currents, and in Sec.~\ref{sec7} we present conclusions
which can be drawn from our study. In the companion paper\cite{[Dob00b]},
we discuss physical aspects of the symmetry-breaking schemes
pertaining to the point groups in question.


\section{Symmetry operators}
\label{sec2}

The point groups of interest in this paper can be introduced in two
ways. The first one consists in defining an abstract point group by
giving its table of multiplications, and then classifying the states
and operators in the fermion Fock space according to the relevant
irreducible representations (irreps). The second one, which we follow
below, is more intuitive, and consist in defining the symmetry
operators in the Fock space first, and then identifying their
multiplication tables and the corresponding group structures.

\subsection{Fock-space representations}
\label{subsec2a}

It will be convenient to use the Cartesian representation of the
symmetry operators. Let $\hatI_k$ for $k=x,y,z$  denote the
Cartesian components of the total angular momentum operator
(generators of the group of rotations).
In the coordinate-space representation these operators read ($\hbar$=1)
\be\label{tam}
\hatI_k \equiv \sum_{n=1}^A
                {\hat{j}}_k^{(n)}
        = \hatL_k + \hat{\Sigma}_k
        = \sum_{n=1}^A
          \left({\hat{l}}_k^{(n)} + \tdemi\hatsi_k^{(n)}\right),
\ee
where ${\hat{j}}_k^{(n)}$, ${\hat{l}}_k^{(n)}$, and
$\tdemi\hatsi_k^{(n)}$ are operators of the total, orbital, and
intrinsic angular momenta, respectively, of particle number $n$. By
definition, operators (\ref{tam}) act in the Hilbert space
${\mathcal{H}}_A$ of $A$-particle states, and the number of particles
$A$ appears explicitly in their definitions.

One can use another representations of $\hatI_k$,
the so-called second-quantized, or Fock-space form,
\be\label{tama}
\hatI_k = \sum_{\mu\nu} \langle\mu|\hat{j}_{k}|\nu\rangle a^+_\mu a_\nu,
\ee
where $\langle\mu|\hat{j}_{k}|\nu\rangle$ are the matrix elements of the angular-momentum
operators in the single-particle basis defined by the fermion
creation and annihilation operators $a^+_\mu$ and $a_\nu$. Operators
$\hatI_k$ in the form of Eq.~(\ref{tama}) do not explicitly depend on
$A$, and act simultaneously in all the $A$-particle spaces, i.e., they
act in the Fock space ${\mathcal{H}}$,
\be\label{Fock}
    {\mathcal{H}} \equiv {\mathcal{H}}_0\oplus{\mathcal{H}}_1\oplus\ldots
                     \oplus{\mathcal{H}}_A\oplus\ldots.
\ee
In each subspace ${\mathcal{H}}_A$, operators (\ref{tama}) are equal
to (\ref{tam}). Since both act in different domains,  one should, in
principle, denote them with different symbols. However, one usually
understands definition (\ref{tam}) as a prescription to construct
$\hatI_k$ for all values of $A$ simultaneously (adjoined by
$\hatI_k$$\equiv$0 for $A$=0). With this extension, operators
(\ref{tam}) and (\ref{tama}) are equal. In this section we understand
that all operators act in the Fock space (\ref{Fock}), while the
corresponding definitions are given in the coordinate-space
representation.

We introduce three standard
transformations
of rotation around three mutually perpendicular axes, ${\mathcal{O}}x$,
${\mathcal{O}}y$ and ${\mathcal{O}}z$, through the angles of $\pi$
as
   \be\label{Eq201}
      \hatR{k} \equiv e^{-i\pi\hatI_k}
               = \bigotimes_{n=1}^A e^{-i\pi{\hat{j}}_k^{(n)}}.
   \ee
Similarly, we introduce three operators of reflection
in planes $yz$, $zx$, and $xy$, for $k=x,y,z$,
respectively, which can be written as
   \be\label{Eq202}
      \hatS{k} \equiv \hatP\hatR{k} ,
   \ee
where the inversion operator is denoted by $\hatP$.
The order of operators in Eq.~(\ref{Eq202}) is unimportant because
   \be\label{Eq202a}
      \lbrack\hatP, \hatR{k}\rbrack=0.
   \ee
Finally, the (antilinear) time-reversal operator in the
coordinate-space representation is defined as \cite{[Mes62]}:
   \be\label{eq206}
      \hatT \equiv \bigotimes_{n=1}^A\left(-i\hatsi_y^{(n)}\right) \hatK ,
   \ee
where $\hatK$ is the complex conjugation operator associated with the
coordinate representation.

In what follows it will be convenient to denote with separate
symbols the products of $\hatT$ with $\hatP$, $\hatR{k}$, and
$\hatS{k}$ \cite{transposition}, i.e., the seven additional
(apart from $\hatT$ itself) antilinear
operators read
   \bnll{Eq204}
      \hatPT    &\equiv& \hatP   \hatT ,       \label{Eq204a}\\
      \hatRT{k} &\equiv& \hatR{k}\hatT ,       \label{Eq204b}\\
      \hatST{k} &\equiv& \hatS{k}\hatT .       \label{Eq204c}
   \enll
The order of multiplications in the
above definitions is irrelevant since
   \be\label{Eq205}
      \lbrack\hatP,   \hatT\rbrack
    = \lbrack\hatR{k},\hatT\rbrack
    = \lbrack\hatS{k},\hatT\rbrack=0.
   \ee

In nuclear physics applications the linear operators $\hatP$,
$\hatR{k}$, and $\hatS{k}$ are usually referred to as inversion,
signature, and simplex. The antilinear operators  $\hatPT$,
$\hatRT{k}$, and $\hatST{k}$ will be called $T$-inversion,
$T$-signature, and $T$-simplex, respectively.

For completeness, yet two another operators must be added to the
above symmetry operators. One of them is, of course, the identity
operator, $\hatE$, which can be treated as the rotation through angle
equal to 0 or $4\pi$ about an arbitrary axis. The second one is the
rotation through angle $2\pi$ about an arbitrary axis, i.e.,
\be\label{2pi}
\barE \equiv e^{-i2\pi\hat{I}_k}
=e^{-i2\pi\hatL_k}\bigotimes_{n=1}^A\left(-\hatsi_0^{(n)}\right)
= (-1)^A\hatE
\ee
where $\hatsi_0$ is the unity $2\times 2$ matrix. We see that
only for even systems $\barE$ is equal to identity, while for odd
systems it is equal to the minus identity. We should keep in mind, that in
the group theory there is no such a notion as a change of sign.
Operators like $(-1)^A$ may appear in representations, like here they
do appear in the Fock-space representation, however, one cannot use
them when defining the group structures in Secs. \ref{single} and
\ref{double} below.

To investigate multiplication rules of the symmetry operators
introduced above one explicitly calculates products of them. For
example, the products of two signatures are
\bn
\hatR{k}\hatR{m}&=&e^{-i\pi\hatL_k}e^{-i\pi\hatL_m}\bigotimes_{n=1}^A
\left(-\hatsi_k^{(n)}\hatsi_m^{(n)}\right)  ,\label{RR}
\en
and the square of the time reversal reads
\be\label{T2}
{\hatT}^2= \bigotimes_{n=1}^A\left(-({\hatsi}_y^{(n)})^2\right)=\barE.
\ee
It is obvious that these results depend on whether $A$ is even or odd.

Therefore, in what follows we introduce notation which explicates
whether the operators act in even or odd fermion spaces,
${\mathcal{H}}_+$ or ${\mathcal{H}}_-$,
\bnll{Fock2}
    {\mathcal{H}}_+ &\equiv& {\mathcal{H}}_0\oplus{\mathcal{H}}_2\oplus\ldots
                       \oplus{\mathcal{H}}_{A=2p}\oplus\ldots,
                                                     \label{Fock2a} \\
    {\mathcal{H}}_- &\equiv& {\mathcal{H}}_1\oplus{\mathcal{H}}_3\oplus\ldots
                       \oplus{\mathcal{H}}_{A=2p+1}\oplus\ldots.
                                                     \label{Fock2b}
\enll
Any Fock-space operator
$\hatU$:${\mathcal{H}}$$\longrightarrow$${\mathcal{H}}$,
which conserves the particle number, is split into two parts
with different domains, i.e.,
\be\label{split}
\hatU = {\hatBU} + {\hatMU},
\ee
where the bold symbols denote operators which act in the even-$A$ spaces,
while the script symbols denote those acting in the odd-$A$ spaces, i.e.,
\bnll{split2}
     {\hatBU}&:&{\mathcal{H}}_+\longrightarrow{\mathcal{H}}_+, \label{split2a} \\
     {\hatMU}&:&{\mathcal{H}}_-\longrightarrow{\mathcal{H}}_-. \label{split2b}
\enll
With these definitions we are now in a position to investigate the
group structures appearing for the introduced operators.

\subsection{Single group {\DT} for even systems}\label{single}

{}For even fermion numbers we consider the Fock-space operators
defined in Sec.~\ref{subsec2a}, and restrict them to ${\mathcal{H}}_+$,
Eqs.~(\ref{split}) and (\ref{split2a}).
Then, the complete multiplication table reads
\bnll{Eq206}
   \hatBR{k}^2=\hatBS{k}^2={\hatBT}^2
&=&\hatBE , \label{Eq206e}\\
   \left({\hatBRT{k}}\right)^2=\left({\hatBST{k}}\right)^2= {\hatBP}^2
&=&\hatBE , \label{Eq206a}\\
   \hatBR{k}\hatBS{k}=\hatBS{k}\hatBR{k}
&=&\hatBP , \label{Eq206ba}\\
   \hatBRT{k}\hatBST{k}=\hatBST{k}\hatBRT{k}
&=&\hatBP , \label{Eq206bb}\\
   \hatBR{k}\hatBRT{k}=\hatBRT{k}\hatBR{k}=\hatBS{k}\hatBST{k}=\hatBST{k}\hatBS{k}
&=&\hatBT , \label{Eq206c}\\
   \hatBR{k}\hatBST{k}=\hatBST{k}\hatBR{k}=\hatBRT{k}\hatBS{k}=\hatBS{k}\hatBRT{k}
&=&\hatBPT , \label{Eq206d}
\enll
for $k=x,y,z$, and
\bnll{Eq207}
     \hatBR {k}\hatBR {l} = \hatBS {k}\hatBS {l}
   = \hatBRT{k}\hatBRT{l} = \hatBST{k}\hatBST{l} &=& \hatBR {m}, \label{Eq207a}\\
     \hatBR {k}\hatBS {l} = \hatBS {k}\hatBR {l}
   = \hatBRT{k}\hatBST{l} = \hatBST{k}\hatBRT{l} &=& \hatBS {m}, \label{Eq207b}\\
     \hatBR {k}\hatBRT{l} = \hatBRT{k}\hatBR {l}
   = \hatBS {k}\hatBST{l} = \hatBST{k}\hatBS {l} &=& \hatBRT{m}, \label{Eq207c}\\
     \hatBRT{k}\hatBS {l} = \hatBS {k}\hatBRT{l}
   = \hatBR {k}\hatBST{l} = \hatBST{k}\hatBR {l} &=& \hatBST{m}, \label{Eq207d}
\enll
for $(k,l,m)$ being an {\em arbitrary} permutation of $(x,y,z)$.

We see that the 16 operators acting in the even-$A$ fermion
spaces constitute the Abelian single group which we denote by {\DT},
   \be\label{Eq208}
      \mbox{\DT}:
      \quad
      \{ \hatBE, \hatBP,  \hatBR{k},  \hatBS{k},
         \hatBT, \hatBPT, \hatBRT{k}, \hatBST{k} \},
   \ee
for $k=x,y,z$. The half of the elements in Eq.~(\ref{Eq208}) are linear
operators and the other half are antilinear.

It follows from the multiplication table of the {\DT} operators,
Eqs.~(\ref{Eq206}) and (\ref{Eq207}),
that the whole group can be generated by its four elements only.
These elements are called the group generators.
Various possibilities of choosing the generators
are discussed in Ref.\cite{[Dob00b]}; here we only mention that,
e.g., the subset \{$\hatBT$, $\hatBP$, $\hatBR{x}$, $\hatBR{y}$\}
can be used to obtain all the operators that belong to {\DT}.

The {\DT} group has two important subgroups, the eight-element Abelian
group {\Dh} composed of all the linear {\DT} operators,
   \be\label{Eq203}
      \mbox{\Dh}: \quad \{ \hatBE, \hatBP, \hatBR{k}, \hatBS{k} \},
   \ee
and the four-element Abelian group {\Dd},
   \be\label{Eq203a}
      \mbox{\Dd}: \quad \{ \hatBE, \hatBR{k} \}.
   \ee

Obviously, the {\Dh}  subgroup of {\DT}, being an Abelian group of
order eight, has eight one-dimensional irreps. These can be labelled by
eigenvalues equal to either +1 or $-$1 of three of its generators,
say, $\hatBP$, $\hatBR{x}$ and $\hatBR{y}$.

We introduce names for the (one-dimensional) bases of the associated
irreps according to the following convention. First, a basis is
called {\em invariant} if it remains unchanged (belongs to eigenvalue
+1) under all three signature operators $\hatBR{k}$, and it is called
either $x$-, or $y$-, or $z$-{\em covariant} if it transforms under
the signature operators like the $x$, or $y$, or $z$ coordinates,
respectively. Secondly, prefix {\em pseudo} is added for bases which
are odd, i.e., belong to eigenvalue $-$1 with respect to the
inversion $\hatBP$.

Since $\hatBT$ is an antilinear operator and also an involutive operator
[i.e., its square is equal to
identity, Eq.~(\ref{Eq206e})], we can always choose the phases of all the basis
states so that they belong to the eigenvalue $T=$+1 of
$\hatBT$\cite{[Mes62]} (see Sec.~\ref{sec4a} below).
In this way we construct eight irreducible one-dimensional
corepresentations (ircoreps) of {\DT}, all being even with respect
to the time reversal. After Wigner \cite{[Wig59]}, we call the
representations of a group containing antilinear operators {\em
corepresentations} to emphasize the fact that they are {\em not} the
representations in the usual sense (see Appendix for details).
By a suitable change of phases of the basis states we can obtain
another set of eight ircoreps of {\DT}, all of them odd (i.e., belonging
to the eigenvalue $T$=$-1$) with respect to the time reversal. We use
prefix {\em anti} to name these time-odd ircoreps. Obviously,
time-even and time-odd ircoreps are pairwise equivalent
(cf.~Ref.\cite{[Kos63]}).

Note that all operators acting in the even fermion spaces
${\mathcal{H}}_+$ can also be classified according to the same set of
sixteen ircoreps of {\DT}. All these ircoreps are listed in Table
\ref{ldt1} together with explicit transformation properties of
several examples of one-particle operators belonging to each ircorep.

\subsection{Double group {\DTD} for odd systems}\label{double}

{}For odd fermion numbers we consider the Fock-space operators
defined in Sec.~\ref{subsec2a}, and restrict them to ${\mathcal{H}}_-$,
Eqs.~(\ref{split}) and (\ref{split2b}).
Since operator $\barME$ [odd-fermion-number part of $\barE$ of
Eq.~(\ref{2pi})] is now an independent group element, additional {\em
partner} operators should be introduced in order to construct the
double group {\DTD}, i.e,
\be\label{Eq211}
   \barMP =\barME\hatMP, \quad
   \barMT =\barME\hatMT, \quad
   \barMPT=\barME\hatMPT,
\ee
and
\be
   \barMR{k}  = \barME \hatMR{k}  ,~   \barMS{k}  = \barME \hatMS{k} , ~
   \barMRT{k} = \barME \hatMRT{k} ,~   \barMST{k} = \barME \hatMST{k} ,
         \label{Eq216}
\ee
for $k$=$x,y,z$. Now
the group multiplication table reads
\bnll{Eq217}
   \hatMR{k}^2=\hatMS{k}^2={\hatMT}^2
&=&\barME , \label{Eq217e}\\
   \left({\hatMRT{k}}\right)^2=\left({\hatMST{k}}\right)^2= {\hatMP}^2
&=&\hatME , \label{Eq217a}\\
   \hatMR{k}\hatMS{k}=\hatMS{k}\hatMR{k}
&=&\barMP , \label{Eq217ba}\\
   \hatMRT{k}\hatMST{k}=\hatMST{k}\hatMRT{k}
&=&\hatMP , \label{Eq217bb}\\
   \hatMR{k}\hatMRT{k}=\hatMRT{k}\hatMR{k}=\hatMS{k}\hatMST{k}=\hatMST{k}\hatMS{k}
&=&\barMT , \label{Eq217c}\\
   \hatMR{k}\hatMST{k}=\hatMST{k}\hatMR{k}=\hatMRT{k}\hatMS{k}=\hatMS{k}\hatMRT{k}
&=&\barMPT , \label{Eq217d}
\enll
for $k=x,y,z$, and
\bnll{Eq218}
     \hatMR {k}\hatMR {l} = \hatMS {k}\hatMS {l}
   = \hatMRT{k}\hatMRT{l} = \hatMST{k}\hatMST{l} &=& \barMR {m}, \label{Eq218a}\\
     \hatMR {k}\hatMS {l} = \hatMS {k}\hatMR {l}
   = \hatMRT{k}\hatMST{l} = \hatMST{k}\hatMRT{l} &=& \barMS {m}, \label{Eq218b}\\
     \hatMR {k}\hatMRT{l} = \hatMRT{k}\hatMR {l}
   = \hatMS {k}\hatMST{l} = \hatMST{k}\hatMS {l} &=& \barMRT{m}, \label{Eq218c}\\
     \hatMRT{k}\hatMS {l} = \hatMS {k}\hatMRT{l}
   = \hatMR {k}\hatMST{l} = \hatMST{k}\hatMR {l} &=& \barMST{m}, \label{Eq218d}
\enll
for $(k,l,m)$ being an {\em odd} permutation of $(x,y,z)$, while
relations identical to (\ref{Eq207}) hold for an {\em even} permutation
of $(x,y,z)$.
After multiplying relations  (\ref{Eq217}) and (\ref{Eq218}) by $\barME$
once or twice, one can easily obtain the remaining elements of the
multiplication table, i.e., those which pertain to products involving
one or two partner
operators (\ref{Eq216}).

The {\DTD} group is thus composed of 32 operators:
\bn
   \mbox{\DTD} : \quad
  \{&\hatME&, \hatMP, \hatMT, \hatMPT,
        \hatMR{k}, \hatMS{k}, \hatMRT{k}, \hatMST{k},   \label{Eq219} \\
    &\barME&, \barMP, \barMT, \barMPT,
        \barMR{k}, \barMS{k}, \barMRT{k}, \barMST{k} \}.       \nonumber
\en
One can see that this double
group is not Abelian, because relations (\ref{Eq218}) now do depend on
whether
the permutation $(k,l,m)$ of $(x,y,z)$ is even or odd, whereas
for the single group, relations (\ref{Eq207}) are independent of that.

One may note that the Fock-space operators $\hatU$ of
Sec.~\ref{subsec2a} and the odd-fermion-space operators $\hatMU$ of
Eqs.~(\ref{split}) and (\ref{split2b}) obey exactly the same
multiplication rules of the double {\DTD} group. Therefore, one
might, in principle, consider only the double group {\DTD} and
refrain from studying the group structures in even and odd spaces
separately. However, at the level of representations, one would then
have been deprived of important properties of operators like
$\hatBT^2$=$\hatBE$ or $\hatMT^2$=$-\hatME$ (see Sec.~\ref{sec4a}),
neither of which holds in
the whole Fock space, cf.~Eqs.~(\ref{T2}) and (\ref{2pi}).

Since the squares of the time reversal, signatures, and simplexes,
Eq.~(\ref{Eq217e}), are equal to $\barE$, the whole double group  {\DTD}
can be generated by the same operators which generate the
single group in even systems, Sec.~\ref{single}.
So the double group also needs four generators; for instance,
the set of four elements, $\hatMT$, $\hatMP$, $\hatMR{x}$, and
$\hatMR{y}$,
can be used to obtain the entire double group of the {\DTD} operators.

The linear operators of the {\DTD} double group form the sixteen-element
double group {\DD},
   \be\label{Eq214}
       \mbox{\DD} : \quad
      \{ \hatME, \hatMP, \hatMR{k}, \hatMS{k},
         \barME, \barMP, \barMR{k}, \barMS{k} \}.
   \ee
This group has 10 equivalence classes
(cf. Refs. [2,3,5]). There are 6 classes composed of 2 elements each,
i.e.,
$\{\hatMR{k}, \barMR{k}\}$ and $\{\hatMS{k},\barMS{k} \}$ for $k=x,y,z$,
while the remaining elements:
$\{\hatME\}$, $\{\barME\}$, $\{\hatMP\}$ and $\{\barMP\}$ form 4
one-element classes by themselves.
The group is not Abelian and possesses, apart from the 8 one-dimensional
irreps already known for the single group, another 2 two-dimensional spinor
irreps.
The spinor irreps can be labelled by the parities (the eigenvalues $\pi$=+1
or $-$1 of the inversion operators $\hatMP$), see Appendix.

The time reversal $\hatMT$ is an antilinear and also an antiinvolutive operator
[i.e., its representations give the minus identity when squared,
Eq.~(\ref{Eq217e})],
and therefore it cannot be diagonalized \cite{[Mes62]}
(see Sec.~\ref{sec4a} below),
and used for labelling the {\DTD} ircoreps. A Hermitian antilinear
involutive operator,
i.e., either a $T-$signature or a $T-$simplex [see Eq.~(\ref{Eq217a})]
should be chosen to serve this purpose. For instance, a pair of commuting
Hermitian operators $\hatMP$ and $\hatMRT{y}$ can be used to label the
{\DTD} ircoreps.
As for the single group {\DT}, ircoreps being
either even or odd with respect to $\hatMRT{y}$  can be obtained one from
another
by a suitable change of phase, and are therefore equivalent. In  analogy
to the one-dimensional ircoreps, the bases of spinor ircoreps belonging to
pairs of
eigenvalues of $\{\hatMP,\hatMRT{y} \}$ equal to $\{+1,+1\}$, $\{-1,+1\}$,
$\{+1,-1\}$ and $\{-1,-1\}$ can be called the {\em spinor}, {\em pseudospinor},
{\em antispinor} and {\em antipseudospinor}
bases, respectively.

Note that only spinor ircoreps of {\DTD} appear in the classification
of states of systems with odd numbers of fermions. However, the
operators acting in ${\mathcal{H}}_-$ can all be classified according
to the one-dimensional ircoreps of {\DTD}, similarly as operators
acting in ${\mathcal{H}}_+$ can all be classified according to the
corresponding one-dimensional ircoreps of {\DT}. (This is completely
analogous to the fact that fermion-number conserving operators can
carry only integer angular momenta, i.e., they are integer-rank
tensors.) Therefore, whenever we consider the action of the {\DT} or
{\DTD} operators on fermion states we always specify whether they act
in even ${\mathcal{H}}_+$ or odd ${\mathcal{H}}_-$ spaces, and use
for them the corresponding notations $\hatBU$ and $\hatMU$ of
Eqs.~(\ref{split2a}) and (\ref{split2b}). On the other hand, whenever we
consider transformation properties $\hatU^\dagger\hatO\hatU$ of
operators $\hatO$ with respect to the {\DTD} group, we do not make
this distinction, and use for them notation $\hatU$ of
Sec.~\ref{subsec2a}.

\subsection{Cartesian harmonic oscillator basis}\label{hoba}

One often uses the Cartesian harmonic oscillator (HO) basis
to solve the self-consistent equations when neither spherical
nor axial symmetry is assumed, see, e.g., Refs.~\cite{[Gir83],[Dob97]}.
The Cartesian HO states are identified by
the numbers of oscillator quanta, $n_x$, $n_y$, and $n_z$, in the three
Cartesian directions,
and by the spin projection $s_z$=$\pm\demi$ on the $z$ axis.
For the standard HO phase convention, this basis is real,
$\hatMK\hoket{s_z}=\hoket{s_z}$, where according to our standard
convention the script symbol $\hatMK$ denotes the coordinate-space
complex-conjugation operator acting in the odd-fermion-number
space ${\mathcal{H}}_-$, see Sec.~\ref{subsec2a}.

{}For the HO states the following relations hold \cite{Var-Bohr}:
   \bnll{eq209}
      \hatMP   \hoket{s_z} &=& \phantom{i}(-1)^{n_x+n_y+n_z}\hoket{s_z} ,
                                                               \label{eq209a}\\
      \hatMT   \hoket{s_z} &=& \phantom{i}(-1)^{\demi-s_z}\hoket{-s_z} ,
                                                               \label{eq209b}\\
      \hatMR{x}\hoket{s_z} &=& i(-1)^{n_y+n_z+1}\hoket{-s_z} ,
                                                               \label{eq209c}\\
      \hatMR{y}\hoket{s_z} \!&=&\!\! \phantom{i}(-1)^{n_x+n_z+\demi-s_z}\hoket{-s_z} ,
                                                               \label{eq209d}\\
      \hatMR{z}\hoket{s_z} &=& i(-1)^{n_x+n_y+\demi+s_z}\hoket{s_z} ,
                                                               \label{eq209e}\\
      \barME   \hoket{s_z} &=& - \hoket{s_z} ,
                                                               \label{eq209f}
   \enll
{}from where one can find similar equations for all the remaining
operators of group {\DTD}. Since the HO Hamiltonian
is symmetric under {\DTD}, its eigenstates can be
classified according to the ircoreps of {\DTD}. It is easily seen that
the HO states $\hoket{s_z}$ form bases of the spinor, pseudospinor,
antispinor and antipseudospinor ircoreps for
$\{N$=$n_x$+$n_y$+$n_z$, $N_y$=$n_x$+$n_z\}$ being \{even, odd\},
\{odd, odd\}, \{even, even\} and \{odd, even\}, respectively (see Appendix). The
entire HO basis would have belonged to the spinor and pseudospinor
ircoreps only, if the basis states and phase convention were chosen differently,
see Ref.\cite{[Dob00b]}.

\subsection{Properties of the {\DT} and {\DTD} operators}
\label{sec4a}

In this section we recall properties of the {\DT} and {\DTD} operators
when they are represented in the fermion Fock space.
Within representations,
apart from the corresponding multiplication tables,
Eqs.~(\ref{Eq206})--(\ref{Eq207}) and (\ref{Eq217})--(\ref{Eq218}),
these operators are characterized
by their hermitian-conjugation properties. Since
all the Fock-space representations of the {\DT} and {\DTD} operators are unitary,
they are hermitian
or antihermitian depending on whether they are involutive
or antiinvolutive, respectively. Properties
of these operators are very different depending on whether they
are linear or antilinear. These characteristics are summarized
in Tables \ref{tab9} and \ref{tab8}, where the {\DT} and {\DTD}
operators are split into two or four subsets, respectively. Below we
review the properties of operators in each such subset.

{\vbox{
\begin{table}[h]
\caption[TT]{Properties of the {\DT} operators $\hatBU$ in even fermion spaces.
\label{tab9}}
\begin{center}
\begin{tabular}{l|ll}
                           & Linear              & Antilinear                   \\
\hline
Hermitian $(\hatBU^2$=$\hatBE)$
                           & $\hatBP$,     $\hatBR{k}$,  $\hatBS{k}$
                           & $\hatBRT{k}$, $\hatBST{k}$, $\hatBT$,    $\hatBPT$ \\
\end{tabular}
\end{center}
\end{table}
}}

{\vbox{
\begin{table}[h]
\caption[TT]{Properties of the {\DTD} operators $\hatMU$ in odd fermion spaces.
\label{tab8}}
\begin{center}
\begin{tabular}{l|ll}
                                     & Linear                   & Antilinear                 \\
\hline
Hermitian $(\hatMU^2$=$\hatME)$      & $\hatMP$                 & $\hatMRT{k}$, $\hatMST{k}$ \\
Antihermitian $(\hatMU^2$=$-\hatME)$ & $\hatMR{k}$, $\hatS{k}$  & $\hatMT$, $\hatMPT$        \\
\end{tabular}
\end{center}
\end{table}
}}

{}For each linear {\DT} or {\DTD} operator one can attribute quantum numbers to
fermion states. These quantum numbers can be equal to $\pm1$ or $\pm{i}$
for hermitian (involutive) or antihermitian (antiinvolutive) operators, respectively.
Therefore, the parity operators $\hatBP$ or $\hatMP$ give the
parity quantum numbers, $\pi$=$\pm1$, the signature operators
$\hatBR{k}$ give the signature quantum numbers, $r$=$\pm1$, in even systems and
the signature operators
$\hatMR{k}$ give
$r$=$\pm{i}$ in odd systems. Likewise, the simplex operators $\hatBS{k}$
and  $\hatMS{k}$
give the
simplex quantum numbers, $s$=$\pm1$ and $s$=$\pm{i}$, respectively.

Antilinear operators do not give good quantum numbers, and their role is very
different,
depending on whether they are hermitian or antihermitian,
Tables \ref{tab9} and \ref{tab8}.

{}For each hermitian antilinear {\DT} or {\DTD} operator, i.e.,
for $\hatBRT{k}$, $\hatBST{k}$, $\hatBT$, $\hatBPT$,
$\hatMRT{k}$, or $\hatMST{k}$
one can find a basis consisting solely of its eigenstates with
the common eigenvalue equal to 1 \cite{[Mes62]}.  Indeed, if
state $|\Psi\rangle$ is an eigenstate of, e.g., $\hatRT{k}$,
the corresponding eigenvalue must be a phase, i.e.,
$\hatRT{k}|\Psi\rangle$=$e^{2i\phi}|\Psi\rangle$.  In such a
case, state $|\Psi'\rangle$=$e^{i\phi}|\Psi\rangle$ is an
eigenstate of $\hatRT{k}$ with eigenvalue 1.  This
demonstrates explicitly that properties of eigenstates of
$\hatRT{k}$ are, of course, phase-dependent.  In the case when
state $|\Psi\rangle$ is not an eigenstate of $\hatRT{k}$,
one can transform the two linearly-independent states
$|\Psi\rangle$ and $\hatRT{k}|\Psi\rangle$ into eigenstates of
$\hatRT{k}$ with eigenvalue 1 by symmetrization and antisymmetrization
of the two:
   \bnll{eq302}
   |\Psi_s\rangle =& |\Psi\rangle +& \hatRT{k}|\Psi\rangle,  \\
   |\Psi_a\rangle =&i|\Psi\rangle -&i\hatRT{k}|\Psi\rangle,
   \enll
which also requires a specific phase.  Therefore,
phase-convention properties of states are essential for a
discussion of bases of eigenstates of the hermitian antilinear
operators, and in Ref.\cite{[Dob00b]} a special discussion
is devoted to this problem.

One should also remember,
that only linear combinations of basis states with {\em real} coefficients
remain eigenstates of any hermitian antilinear operator.
This is in contrast to properties of linear operators,
for which a linear combination of eigenstates, corresponding
to the same eigenvalue, with {\em arbitrary}
coefficients, is also an eigenstate with the same eigenvalue.

Very special properties characterize the antihermitian antilinear
operators. Within the {\DT} or {\DTD} groups only the
$\hatMT$ and $\hatMPT$ operators in odd systems belong to such a subset
(Table \ref{tab8}). For each antihermitian antilinear {\DTD}
operator the space of fermion states can be
arranged in pairs of orthogonal states $(|\Psi(+)\rangle,|\Psi(-)\rangle)$
\cite{[Mes62]}, such that, for example,
   \be\label{eq305}
   \hatMT |\Psi(\pm)\rangle=\pm |\Psi(\mp)\rangle ,
   \ee
or
   \be\label{eq306}
   \hatMPT |\Psi(\pm)\rangle=\pm |\Psi(\mp)\rangle .
   \ee
Therefore, the $\hatMT$ and $\hatMPT$ operators cannot be diagonalized.
In particular, there is no odd fermion state
which would be invariant with respect to  $\hatMT$ or $\hatMPT$.


\section{Symmetries of local densities}
\label{sec6}

Suppose that the Fock-space operator
$\hatBU$ or $\hatMU$, belonging to {\DT} or {\DTD}, respectively,
represents a symmetry conserved by
a mean-field many-particle state $|\Psi_+\rangle$ or $|\Psi_-\rangle$,
in even or odd fermion spaces, i.e.,
\bnll{ld1a}
    {\hatBU}|\Psi_+\rangle = u|\Psi_+\rangle \label{ld1aa} , \\
    {\hatMU}|\Psi_-\rangle = u|\Psi_-\rangle \label{ld1ab} .
\enll
As discussed in Sec.~\ref{sec4a}, eigenvalue $u$ can be equal to
$\pm1$ or $\pm{i}$, and moreover, in odd fermion systems $\hatU$
cannot be equal to either $\hatMT$ or $\hatMPT$, i.e, neither the
time reversal nor the product of inversion and time reversal can be a
conserved symmetry in odd systems. According to conventions
introduced in Sec.~\ref{double}, in
odd systems the hat always denotes one of the {\DTD} operators
introduced in Sec.~(\ref{subsec2a}), and
not one of their partners [Eqs.~(\ref{Eq211}) and (\ref{Eq216})].
Of course,
if $\hatMU$ is a symmetry of $|\Psi_-\rangle$ then $\barMU$ is a symmetry
as well, so from the point of view of conserved symmetries, any extra
study of partner operators is unnecessary.

\twocolumn[
\mediumtext
{\begin{table}[h]
{\caption[TT]{
Symmetry properties of space-spin one-particle operators $\hatO$ belonging to
different one\--dimensional ircoreps of the {\DT} and {\DTD} groups. The
first column gives names of different ircoreps, the second column lists
examples of operators $\hatO$, and the remaining
columns give signs in the expression
$\hatU^\dagger\hatO\hatU$=$\pm$$\hatO$, for operators $\hatU$
given in the column headers. Note that these results do not depend on
whether operators $\hatU$ act in even or odd spaces,
and therefore the Fock-space notation is used for them.
\label{ldt1}}}
\begin{center}
\begin{tabular}{l|l|cccc}
$\mbox{Ircorep} $&$
\mbox{Space-spin one-particle operators $\hatO$}
              $&$ \hatT      $&$\hatP   $&$\hatR{x}  $&$\hatR{y}$ \\
\hline
${\mbox{{\inv}s}}  $&$
x^2,y^2,z^2,\nabla_x^2,\nabla_y^2,\nabla_z^2;
xi\nabla_y\hatsi_z, yi\nabla_z\hatsi_x, zi\nabla_x\hatsi_y
                               $&$  +   $&$  +   $&$  +   $&$  +   $\\

${\mbox{{\ps}{\inv}s}}  $&$
xyz, xi\nabla_yi\nabla_z, yi\nabla_zi\nabla_x, zi\nabla_xi\nabla_y;
i\nabla_x\hatsi_x,i\nabla_y\hatsi_y,i\nabla_z\hatsi_z
                               $&$  +   $&$  -   $&$  +   $&$  +   $\\

${\mbox{{\ay}{\inv}s}}  $&$
xi\nabla_x,yi\nabla_y,zi\nabla_z;
xy\hatsi_z, yz\hatsi_x, zx\hatsi_y
                               $&$  -   $&$  +   $&$  +   $&$  +   $\\

${\mbox{{\ay}{\ps}{\inv}s}}  $&$
xyi\nabla_z,yzi\nabla_x, zxi\nabla_y;
x\hatsi_x,y\hatsi_y,z\hatsi_z
                               $&$  -   $&$  -   $&$  +   $&$  +   $\\

${\mbox{{\cov{$x$}}s}}  $&$
x;i\nabla_y\hatsi_z, i\nabla_z\hatsi_y
                                $&$  +   $&$ -   $&$  +   $&$  -   $\\

${\mbox{{\cov{$y$}}s}}  $&$
y;i\nabla_x\hatsi_z, i\nabla_z\hatsi_x
                               $&$  +   $&$  -   $&$  -   $&$  +   $\\

${\mbox{{\cov{$z$}}s}}  $&$
z;i\nabla_x\hatsi_y, i\nabla_y\hatsi_x
                               $&$  +   $&$  -   $&$  -   $&$  -   $\\

${\mbox{{\pscov{$x$}}s}}  $&$
yz;xi\nabla_y\hatsi_y, i\nabla_xy\hatsi_y, \hatsi_xyi\nabla_y
                               $&$  +   $&$  +   $&$  +   $&$  -   $\\

${\mbox{{\pscov{$y$}}s}}   $&$
xz;yi\nabla_z\hatsi_z, i\nabla_yz\hatsi_z, \hatsi_yzi\nabla_z
                               $&$  +   $&$  +   $&$  -   $&$  +   $\\

${\mbox{{\pscov{$z$}}s}} $&$
xy;zi\nabla_x\hatsi_x, i\nabla_zx\hatsi_x, \hatsi_zxi\nabla_x
                               $&$  +   $&$  +   $&$  -   $&$  -   $\\

${\mbox{{\acov{$x$}}s}}  $&$
i\nabla_x;y\hatsi_z,z\hatsi_y
                               $&$  -   $&$  -   $&$  +   $&$  -   $\\

${\mbox{{\acov{$y$}}s}}  $&$
i\nabla_y;x\hatsi_z,z\hatsi_x
                               $&$  -   $&$  -   $&$  -   $&$  +   $\\

${\mbox{{\acov{$z$}}s}}  $&$
i\nabla_z;x\hatsi_y,y\hatsi_x
                               $&$  -   $&$  -   $&$  -   $&$  -   $\\

${\mbox{{\apscov{$x$}}s}}  $&$
yi\nabla_z, zi\nabla_y;\hatsi_x
                               $&$  -   $&$  +   $&$  +   $&$  -   $\\

${\mbox{{\apscov{$y$}}s}}  $&$
xi\nabla_z, zi\nabla_x;\hatsi_y
                               $&$  -   $&$  +   $&$  -   $&$  +   $\\

${\mbox{{\apscov{$z$}}s}}  $&$
xi\nabla_y, yi\nabla_x;\hatsi_z
                               $&$  -   $&$  +   $&$  -   $&$  -   $

\end{tabular}
\end{center}
\end{table}}]\narrowtext

Mean-field state $|\Psi\rangle$ can be characterized by the
single-particle density matrix $\rho$ (see Ref.~\cite{[RS80]} for the
definition), for which the symmetry properties (\ref{ld1a}) imply
\be\label{ld1}
                {\hatU}^\dagger\rho\hatU = \rho,
\ee
independently of eigenvalue $u$. (Symmetry property (\ref{ld1}) does
not depend on whether the mean-field state belongs to the even or odd
fermion space, and therefore, we use the Fock-space notation $\hatU$
for the symmetry operators, see definitions in Sec.~\ref{subsec2a} and
discussion in Sec.~\ref{double}.) It then follows that the
single-particle self-consistent Hamiltonian $h[\rho]$ is also
symmetric with respect to operator $\hatU$ \cite{[RS80]}, namely,
\be\label{ld2}
                 {\hatU}^\dagger h[\rho]\hatU = h[\rho].
\ee
Eq.~(\ref{ld2}) implies that if $\varphi$ is a normalized
single-particle eigenfunction
of $h[\rho]$, then
$\hatU\varphi$ is also a normalized eigenfunction, both belonging
to the same eigenvalue. As a  consequence, it can be shown \cite{[Rip68]}
that the symmetry is preserved during the standard self-consistent
iteration, provided
the entire multiplets of states belonging to the same eigenvalue of $\hatU$ are
either fully occupied, or fully empty.
In such a case Eqs.~(\ref{ld1}) and
(\ref{ld2}) are fulfilled repeatedly in the successive steps of iteration,
and $\hatU$ is a self-consistent symmetry.

Since the one-body density is a fermion-number conserving one-body
operator, it can be classified according to one-dimensional ircoreps
of {\DT} or {\DTD}, and this can be done both in even and odd
systems. This means that either the given operator $\hatU$ is a
conserved symmetry, Eqs.~(\ref{ld1a}) and (\ref{ld1}), and the
density matrix belongs to the given one-dimensional ircorep of the
subgroup generated by $\hatU$, or $\hatU$ is a broken symmetry and
the density matrix has two non-zero components in two different such
one-dimensional ircoreps. It follows that in odd systems the density
matrix has always non-zero components in two ircoreps corresponding
to the time reversal.

This classification procedure is used below to enumerate properties
of the density matrix when one or more {\DT} or {\DTD} operators are
conserved symmetries. Note also, that unlike for the many-body states
$|\Psi\rangle$, one does not have a freedom to change the phase of
the density matrix, because it is a hermitian operator independent of
the phase of the mean-field state it corresponds to. Therefore, if
the density matrix has non-zero components in two ircoreps
corresponding to two different eigenvalues of an antilinear {\DT} or
{\DTD} operator, it cannot be transformed to the form in which it
would have been either even or odd with respect to this operator.

A definite symmetry of the density matrix, Eq.~(\ref{ld1}), implies
certain symmetries for local densities and their derivatives. These
symmetries are discussed and enumerated in the present section.

The spin structure of the density matrix is given by
   \bn
   \rho(\bbox{r}\sigma,\bbox{r}'\sigma')
    & = &\tdemi\rho(\bbox{r},\bbox{r}')\delta_{\sigma\sigma'}
    \nonumber\\
    & + &\tdemi\sum_{k=x,y,z}s_k(\bbox{r},\bbox{r}')
            <\sigma|{\hatsi}_k|\sigma'>, \label{ld3}
   \en
where $\bbox{r}$=$(x,y,z)$ and $\bbox{r}'$=$(x',y',z')$
represent three-di\-men\-sio\-nal position vectors.
When the rotational symmetry is preserved one often refers to
$\rho(\bbox{r},\bbox{r}')$ and $s_k(\bbox{r},\bbox{r}')$ as
the {\em scalar} and {\em vector} densities, respectively.  In
our case, the rotational symmetry is broken, and we will avoid
using these terms.
Instead, we classify the densities
according to the  ircoreps of the {\DT} or {\DTD} group.
As discussed above, for the one-body operators
only the one-dimensional ircoreps are relevant for the classification.
There are 16 characteristic transformation properties
of the bases for one-dimensional ircoreps.
In Table \ref{ldt1} we list all these ircoreps,
illustrated by examples of space-spin operators
of interest, e.g., powers of coordinates, $x$, $y$, $z$ and
gradients, $\nabla_x$, $\nabla_y$, $\nabla_z$.

The Table also
lists explicitly the transformation properties of operators
belonging to every type of symmetry.  For example, the sign
"$-$" which appears in row denoted by "$y$-covariants" and
column denoted by $\hatR{x}$ means that
$\hatR{x}^\dagger y\hatR{x}$=$-y$. It can be easily checked that
 the Pauli matrices, $\hatsi_x$, $\hatsi_y$, $\hatsi_z$ transform
 under the signatures as the $x, y, z$ coordinates, respectively, do not
change under
the inversion, and change their signs under the time reversal.
Therefore, these can be classified as $k$-antipseudocovariants for
$k= x,y,z$, respectively. Spin-dependent
 operators belonging to other ircoreps can also be constructed
{}from the Pauli matrices. Therefore,
 examples of spin-dependent operators are also listed in the Table.
In Table \ref{ldt1} we have introduced the same names for operators  as for
the bases of one-dimensional ircoreps (see Sec. \ref{single}).

Similarly as in Ref.\cite{[Eng75]}, we consider the following local densities:
\begin{itemize}

\item[--] particle and spin densities
   \bnll{ld4}
     \rho(\bbox{r})
     &=& \rho(\bbox{r},\bbox{r}),
                                               \label{ld4a} \\
     s_k(\bbox{r})
     &=& s_k(\bbox{r},\bbox{r}),
                                                \label{ld4b}
   \enll
\item[--] kinetic and spin-kinetic densities
   \bnll{ld5}
     \tau_{kl}(\bbox{r})
     &=&
     \left[\nabla_k{{\nabla}'}_l
     \rho(\bbox{r},\bbox{r}')\right]_{\bbox{r}=\bbox{r}'},
                                                 \label{ld5a} \\
     {T}_{klm}(\bbox{r})
     &=&
     \left[{\nabla}_k{{\nabla}'}_l
     s_m(\bbox{r},\bbox{r}')\right]_{\bbox{r}=\bbox{r}'},
                                                \label{ld5b}
   \enll

\item[--] current and
      spin-current densities
   \bnll{ld6}
     {j}_k(\bbox{r})&=&\frac{1}{2i}
     \left[{\nabla}_k-{{\nabla}'}_k)
     \rho(\bbox{r},\bbox{r}')\right]_{\bbox{r}=\bbox{r}'},
                                                \label{ld6a} \\
     J_{kl}(\bbox{r})&=&\frac{1}{2i}
     \left[(\nabla_k-\nabla'_k)
     s_{l}(\bbox{r},\bbox{r}')\right]_{\bbox{r}=\bbox{r}'},
                                                \label{ld6b}
   \enll
\end{itemize}
where each index $k$, $l$, or $m$ may refer to either of
$x$, $y$, or $z$. It follows
{}from the hermiticity of the density matrix $\rho$  that all the above local
densities are real functions of $\bbox{r}$. Usually
only the traces of kinetic densities,
   \bnll{tauT}
   \tau(\bbox{r}) &=& \sum_k \tau_{kk}(\bbox{r}), \\
   {T}_{m}(\bbox{r}) &=& \sum_k {T}_{kkm}(\bbox{r}),
   \enll
are used in applications.

When operator $\hatU$ represents a conserved symmetry
of the density matrix, Eq.~(\ref{ld1}), the transformation
rules for gradients and spin operators, given in Table \ref{ldt1},
imply definite transformation rules for
the local densities.
These are listed in Table \ref{ldt2}, for all the one-dimensional ircoreps
of {\DT} or {\DTD} as indicated in the first column. In the second column we
show the local densities in forms defined by
Eqs.~(\ref{ld4})--(\ref{ld6}), while the third column gives, when possible,
the local densities in the traditional vector-tensor notation,
e.g.,
   \bnll{vecsT}
   \bbox{s}&=&(s_x,s_y,s_z),\\
     \bbox{T}&=&(T_x,T_y,T_z),
   \enll
   \bnll{vecJ}
             {J}&=&\sum_k J_{kk},\\
   (\tensor{J})_{kl}&=&\demi (J_{kl}+J_{lk})-\frac{1}{3}J\delta_{kl},\\
   (\bbox{J})_k&=& \sum_{lm}\varepsilon_{klm}J_{lm}.
   \enll
Derivatives of densities up to the second order are also
included in the Table.

{}From Table \ref{ldt2} one can read off the symmetry properties of various
densities.  Suppose $d(x,y,z)$ is a generic name of one of the
densities listed in the second or third column, and $\hatU$ is a
generic name of one of the {\DT} or {\DTD} operators listed in the first row.
We use the convention that index $i$ may take any value among $x$,
$y$ or $z$, while indices $k\neq l\neq m$ are arbitrary permutations
of $x$, $y$, and $z$.  If $\hatU$ represents a conserved symmetry,
one has the following symmetry rule for the density $d(x,y,z)$:
   \be\label{consym}
    d(\epsilon_xx,\epsilon_yy,\epsilon_zz)=\epsilon d(x,y,z),
   \ee
where $\epsilon$ is the sign listed in Table
\ref{ldt2} in the row denoted by $d$ and column denoted by
$\hatU$.  Signs $(\epsilon_x,\epsilon_y,\epsilon_z)$ are
given in the last row of Table \ref{ldt1}, and pertain to
two {\DT} or {\DTD} operators ({\em viz.} $\hatU$ and $\hatUT$)
in two adjacent columns.  These latter
signs give changes of coordinates $(x,y,z)$ under the action
of $\hatU$.  As the time reversal does not affect spatial
coordinates, these signs are the same for any pair of
operators $\hatU$ and $\hatU^T$.  One generic Table of signs
determines, therefore, symmetry properties of any local
density for any of the {\DT} or {\DTD} symmetries being preserved.

{}For example, symmetry properties of density $J_{xy}$ can
be found by using indices $l$=$x$ and $m$=$y$ (which requires
$k$=$z$) in the row pertaining to $k$-covariants.  For the
conserved $\hatR{z}$=$\hatR{k}$ symmetry we then find in the
corresponding column $\epsilon$=+ and
$\epsilon_x$=$\epsilon_l$=$-$,
$\epsilon_y$=$\epsilon_m$=$-$, and
$\epsilon_z$=$\epsilon_k$=+, which gives
$J_{xy}(-x,-y,z)$=$J_{xy}(x,y,z)$.

\twocolumn[
\widetext
{\begin{table}[h]
{\caption[TT]{
Symmetry properties of various local densities belonging to
different one-dimensional ircoreps of the {\DT} or {\DTD} groups.
For instructions on using the Table see the
text explaining Eq.~(\protect\ref{consym}).
\label{ldt2}}}
\begin{center}
\begin{tabular}{l|l|l|cccccccccccccc}
$\mbox{Ircorep}                $& \multicolumn{2}{c|}{Local densities}
                                &$\hatP      $&$\hatPT
                               $&$\hatR{k}   $&$\hatRT{k}
                               $&$\hatS{k}   $&$\hatST{k}
                               $&$\hatR{l}    $&$\hatRT{l}
                               $&$\hatS{l}    $&$\hatST{l}
                               $&$\hatR{m}    $&$\hatRT{m}
                               $&$\hatS{m}    $&$\hatST{m}$\\
\hline
${\mbox{{\inv}s}}   $&$\rho,\tau_{ii},\nabla^2_i\rho,\nabla_kJ_{lm}
                    $&$\rho,\tau,\Delta\rho,\bbox{\nabla}\cdot\bbox{J}
    $&$+$&$+$&$+$&$+$&$+$&$+$&$+$&$+$&$+$&$+$&$+$&$+$&$+$&$+$\\
${\mbox{{\ps}{\inv}s}}   $&$ J_{ii}
                         $&$ {J}
    $&$-$&$-$&$+$&$+$&$-$&$-$&$+$&$+$&$-$&$-$&$+$&$+$&$-$&$-$\\
${\mbox{{\ay}{\inv}s}}   $&$ T_{klm}, \nabla_ij_i, \nabla_k\nabla_ls_m
                         $&$ \bbox{\nabla}\cdot\bbox{j}
    $&$+$&$-$&$+$&$-$&$+$&$-$&$+$&$-$&$+$&$-$&$+$&$-$&$+$&$-$\\
${\mbox{{\ay}{\ps}{\inv}s}}  $&$ \nabla_is_i
                         $&$ \bbox{\nabla}\cdot\bbox{s}
    $&$-$&$+$&$+$&$-$&$-$&$+$&$+$&$-$&$-$&$+$&$+$&$-$&$-$&$+$\\
${\mbox{{\cov{$k$}}s}}   $&$ J_{lm},\nabla_k\rho
                         $&$ (\bbox{J})_k,(\tensor{J})_{lm},
				 (\bbox{\nabla}\rho)_k
    $&$-$&$-$&$+$&$+$&$-$&$-$&$-$&$-$&$+$&$+$&$-$&$-$&$+$&$+$\\
${\mbox{{\pscov{$k$}}s}} $&$\tau_{lm},\nabla_l\nabla_m\rho,\nabla_kJ_{ii},
                         $&$(\bbox{\nabla}J)_k,
    $&$+$&$+$&$+$&$+$&$+$&$+$&$-$&$-$&$-$&$-$&$-$&$-$&$-$&$-$\\
$                        $&$\nabla_iJ_{ki},\nabla_iJ_{ik}
                         $&$ (\bbox{\nabla}\cdot \tensor{J})_k,
				 (\bbox{\nabla}\times \bbox{J})_k
     $&$ $&$ $&$ $&$ $&$ $&$ $&$ $&$ $&$ $&$ $&$ $&$ $&$ $&$ $\\
${\mbox{{\acov{$k$}}s}}       $&$j_k,\nabla_ls_m
                              $&$(\bbox{j})_k,
					(\bbox{\nabla}$$\times$$\bbox{s})_k
    $&$-$&$+$&$+$&$-$&$-$&$+$&$-$&$+$&$+$&$-$&$-$&$+$&$+$&$-$\\
${\mbox{{\apscov{$k$}}s}}   $&$s_k,T_{iik},T_{kii},\nabla_lj_m,
                            $&$(\bbox{s})_k,(\bbox{T})_{k},
				 (\bbox{\nabla}$$\times$$\bbox{j})_k
    $&$+$&$-$&$+$&$-$&$+$&$-$&$-$&$+$&$-$&$+$&$-$&$+$&$-$&$+$\\
$                           $&$\nabla_k\nabla_is_i,\nabla_i^2s_k
                            $&$(\bbox{\nabla}(\bbox{\nabla}$$\cdot$$\bbox{s}))_k,
                                 (\Delta \bbox{s})_k
     $&$ $&$ $&$ $&$ $&$ $&$ $&$ $&$ $&$ $&$ $&$ $&$ $&$ $&$ $\\
\hline
\multicolumn{3}{r}{$(\epsilon_k,\epsilon_l,\epsilon_m)=$}&
\multicolumn{2}{c} {$(---)$}&
\multicolumn{2}{c} {$(+--)$}&
\multicolumn{2}{c} {$(-++)$}&
\multicolumn{2}{c} {$(-+-)$}&
\multicolumn{2}{c} {$(+-+)$}&
\multicolumn{2}{c} {$(--+)$}&
\multicolumn{2}{c} {$(++-)$}\\
\end{tabular}
\end{center}
\end{table}}]\narrowtext

It is worth noting that symmetry properties (\ref{consym})
which correspond to various {\DT} or {\DTD} operators, are related
to one another only by the corresponding group multiplication rules.
Therefore, a specific
choice of the conserved generators, either for the complete
{\DT} or {\DTD} groups or for any of their subgroups \cite{[Dob00b]},
leads to a specific set of symmetry properties of local
densities.

Symmetry properties
(\ref{consym}) can be used for the purpose of a continuation
of densities from one semi-space into the second semi-space, i.e.,
one can use only space points for, e.g., $x$$\geq$0.
For two symmetry properties (\ref{consym}), coming from two
different symmetry operators (but not from the pair $\hatU$
and $\hatUT$), one can restrict the space to a quarter-space,
where two coordinates have definite signs, e.g., $x$$\geq$0 and $y$$\geq$0.
Finally, three
conserved symmetries allow for a restriction to one eighth
of the full space with all the coordinates having definite signs, e.g.,
$x$$\geq$0, $y$$\geq$0, and $y$$\geq$0.
The time-reversal symmetry does not lead to
restrictions on the space properties of densities, but, when
conserved, gives the vanishing of all the antiinvariant,
antipseudoinvariant, anticovariant and antipseudocovariant
densities, viz., $s_k, j_k, T_{klm}$ for arbitrary $k,l,m$
as well as their derivatives (see Table \ref{ldt2}).  The
possibilities of simultaneously conserving one, two, three,
or four symmetry operators from the {\DT} or {\DTD} groups will be
discussed in Ref.\cite{[Dob00b]}.

Since density matrix $\rho$ and single-particle Hamiltonian
$h[\rho]$ are always simultaneously invariant under any
conserved symmetry $\hatU$, Eqs.~(\ref{ld1}) and
(\ref{ld2}), the discussion above can be repeated for
self-consistent local fields appearing in a local mean-field
Hamiltonian.  Explicit formulas for symmetry properties of
local fields are identical to those listed in Table
\ref{ldt2}, and will not be repeated here.  In applications,
these symmetries appear automatically when the
self-consistent mean fields are calculated in terms of
densities, cf.~Ref~\cite{[Eng75]}.

\section{Symmetries of shapes, currents, and average angular momenta}
\label{sec8}

In this section we discuss properties of average values of various
operators, calculated for the HF many-particle state $|\Psi\rangle$.
In particular, we consider the electromagnetic multipole operators
and the total angular momentum; the quantities which are used to
characterize properties of investigated systems. First of all, we
enumerate transformation properties of these operators under the
{\DT} or {\DTD} operators. Similarly as for the density matrix
(Sec.~\ref{sec6}), the one-body operators discussed in this section
belong to one-dimensional ircoreps of {\DT} or {\DTD}, and therefore,
their properties do not depend on whether the system contains even or
odd number of fermions.

\subsection{Transformation properties of angular momentum and multipole
operators}\label{multop}

The $k$-component of total angular momentum, $\hatI_k$,
transforms obviously as $k$-antipseudocovariant under {\DT} or {\DTD}, and its
transformation rules can be easily read off from Table \ref{ldt1}.

\twocolumn[
\mediumtext
{\begin{table}
{\caption[TT]{
Symmetry properties of electric multipole operators $\hatQ{\lambda\mu}$ with
respect to operators of the {\DT} or {\DTD} groups. The
results of the symmetry operator $\hatU^\dagger\hatQ{\lambda\mu}\hatU$
are given for three spatial directions $k$=$x$, $y$, $z$.
Where applicable, the upper part of the Table
gives expressions in terms of changed signs
of magnetic components, and the lower part gives the equivalent
expressions in terms of the complex conjugation.
\label{tab2}}}
\begin{center}
\begin{tabular}{c|r@{}lr@{}lr@{}lr@{}l}
$k$&  \multicolumn{2}{c}{$\hatR{k}$}
   &  \multicolumn{2}{c}{$\hatRT{k}$}
   &  \multicolumn{2}{c}{$\hatS{k}$}
   &  \multicolumn{2}{c}{$\hatST{k}$}           \\
\hline
$x$&$(-1)^{\lambda}    $&$\hatQ{\lambda,-\mu}
  $&$(-1)^{\lambda+\mu}$&$\hatQ{\lambda  \mu}
  $&$                  $&$\hatQ{\lambda,-\mu}
  $&$(-1)^{\mu}        $&$\hatQ{\lambda  \mu}  $\\
$y$&$(-1)^{\lambda-\mu}$&$\hatQ{\lambda,-\mu}
  $&$(-1)^{\lambda}    $&$\hatQ{\lambda  \mu}
  $&$(-1)^{-\mu}       $&$\hatQ{\lambda,-\mu}
  $&$                  $&$\hatQ{\lambda  \mu}  $\\
$z$&$(-1)^{\mu}        $&$\hatQ{\lambda  \mu}
  $&$                  $&$\hatQ{\lambda,-\mu}
  $&$(-1)^{\lambda+\mu}$&$\hatQ{\lambda  \mu}
  $&$(-1)^{\lambda}    $&$\hatQ{\lambda,-\mu}  $\\
\hline
$x$&$(-1)^{\lambda+\mu}$&$\hatQ{\lambda  \mu}^*
  $&$(-1)^{\lambda+\mu}$&$\hatQ{\lambda  \mu}
  $&$(-1)^{\mu}        $&$\hatQ{\lambda  \mu}^*
  $&$(-1)^{\mu}        $&$\hatQ{\lambda  \mu}  $\\
$y$&$(-1)^{\lambda}    $&$\hatQ{\lambda  \mu}^*
  $&$(-1)^{\lambda}    $&$\hatQ{\lambda  \mu}
  $&$                  $&$\hatQ{\lambda  \mu}^*
  $&$                  $&$\hatQ{\lambda  \mu}  $\\
$z$&$(-1)^{\mu}        $&$\hatQ{\lambda  \mu}
  $&$(-1)^{\mu}        $&$\hatQ{\lambda  \mu}^*
  $&$(-1)^{\lambda+\mu}$&$\hatQ{\lambda  \mu}
  $&$(-1)^{\lambda+\mu}$&$\hatQ{\lambda  \mu}^*$\\
\end{tabular}
\end{center}
\end{table}}]\narrowtext

{}For $\lambda$ even (odd), the electric multipole operators
$\hatQ{\lambda\mu}$ are even (odd), respectively,  under the action
of the inversion, and are all even with respect to the time reversal,
i.e.,
   \bnll{eq541}
        \hatP^{\dagger}\hatQ{\lambda\mu}\hatP
     &=&  (-1)^{\lambda}\hatQ{\lambda\mu}     , \label{eq541a} \\
        \hatT^{\dagger}\hatQ{\lambda\mu}\hatT
     &=&                     \hatQ{\lambda\mu}^* . \label{eq541b}
   \enll%
The magnetic multipole operators
$\hatM{\lambda\mu}$ have opposite transformation properties,
   \bnll{eq541m}
        \hatP^{\dagger}\hatM{\lambda\mu}\hatP
     &=& - (-1)^{\lambda}\hatM{\lambda\mu}     , \label{eq541ma} \\
        \hatT^{\dagger}\hatM{\lambda\mu}\hatT
     &=&                    - \hatM{\lambda\mu}^* . \label{eq541mb}
   \enll%

Table \ref{tab2} gives  transformation properties\cite{[Var88]}
of $\hatQ{\lambda\mu}$ with
respect to operators of the {\DT} or {\DTD} groups, other than $\hatT$ and $\hatP$.
One may note that
the electric multipole operators are invariant with respect to
the $\hatST{y}$ symmetry.  This is of course a consequence of the
standard phase convention for the rotational irreducible
tensor operators\cite{[Var88],phacon},
   \bnll{eq539}
   \label{eq539a}
        \hatQ{\lambda\mu}^*
     &=&  (-1)^{-\mu}\hatQ{\lambda,-\mu} , \\
   \label{eq539b}
        \hatM{\lambda\mu}^*
     &=&  (-1)^{-\mu}\hatM{\lambda,-\mu} ,
   \enll
which ensures that the antilinear operator $\hatST{y}$ acts as an
identity upon any irreducible spherical tensor operator.


\subsection{Average values}\label{mv}

The electric and magnetic moments are defined as
   \bnll{eq701}
        Q_{\lambda\mu} &=& \langle\Psi |\hatQ{\lambda\mu}|\Psi\rangle
            =  \int q_{\lambda\mu}(\bbox{r})\,d^3\bbox{r}
                                              , \label{eq701a} \\
        M_{\lambda\mu}&=&\langle\Psi |\hatM{\lambda\mu}|\Psi\rangle
            =  \int m_{\lambda\mu}(\bbox{r})\,d^3\bbox{r}
                                              , \label{eq701b}
   \enll%
where $|\Psi\rangle$ is a many-body mean-field state, and
$q_{\lambda\mu}(\bbox{r})$ and $m_{\lambda\mu}(\bbox{r})$
are the corresponding moment densities:
   \bnll{eq702}
        q_{\lambda\mu}(\bbox{r}) &=&
                e\rho(\bbox{r})Q_{\lambda\mu}(\bbox{r})
                                              , \label{eq702a} \\
        m_{\lambda\mu}(\bbox{r}) &=&
          \mu_N\sum_{k=x,y,z}\left(
	    g_s{s_k}\nabla_k Q_{\lambda\mu}(\bbox{r})\right.
	    \nonumber \\
         & - & \left.  {\textstyle\frac{2}{\lambda+1}}
                     g_l{j_k}(\bbox{r}\times
                 \bbox{\nabla}Q_{\lambda\mu}(\bbox{r}))_k\right)
                   , \nonumber \\ &&            \label{eq702b}
   \enll%
and $e$, $g_s$, and $g_l$ are the elementary charge, and the spin and orbital
gyromagnetic factors, respectively \cite{[RS80]}.
In definitions (\ref{eq702}), multipole functions\cite{[Var88]}
(solid harmonics) have the standard
form: $Q_{\lambda\mu}(\bbox{r})$=$r^\lambda Y_{\lambda\mu}(\theta,\phi)$.

Similarly, the mean value of the $k$-component of total angular momentum
(in units of $\hbar$) reads
\be\label{mvtam}
I_k = \langle\Psi |\hatI_k|\Psi\rangle
 = \int \left(\varepsilon_{klm}r_lj_{m}(\bbox{r}) + \tdemi s_k(\bbox{r})\right)
d^3\bbox{r}
\ee

We may now combine symmetry properties of densities $\rho$,
$\bbox{s}$, and $\bbox{j}$, Table \ref{ldt1}, with those of multipole
operators, Table \ref{tab2}, to obtain symmetry conditions obeyed by
the electric and magnetic moments, and by the average angular
momenta, for given conserved symmetries of the {\DT} or {\DTD} groups. In doing
so, we have to remember that since the electric multipole operators
are time-even, the corresponding electric moments depend only on the
time-even component of the density matrix, as given in
Eq.~(\ref{eq702a}). This is so irrespective of whether the
time reversal is, or is not a conserved symmetry, or whether the
system contains even or odd number of fermions. Therefore, the
time reversal does not impose any condition on the electric multipole
moments. On the other hand, with the time-reversal symmetry
conserved, which may occur only for even systems, all magnetic
moments and average angular momenta must vanish, because they depend
only on the time-odd component of the density matrix, Eqs.~(\ref{eq702b})
and (\ref{mvtam}).

\twocolumn[
\mediumtext
{\begin{table}
{\caption[TT]{
Conditions fulfilled by the electric and magnetic multipole moments,
$Q_{\lambda\mu}$ and $M_{\lambda\mu}$, for conserved
{\DT} or {\DTD} operators.
Where applicable, the upper part of the Table
gives expressions in terms of changed signs
of magnetic components, and the lower part gives equivalent
expressions in terms of the complex conjugation.
\label{tab3}}}
\begin{center}
\begin{tabular}{c|r@{}c@{}r@{}lr@{}c@{}r@{}l
                  r@{}c@{}r@{}lr@{}c@{}r@{}l}
$k$&  \multicolumn{4}{c}{$\hatR{k}$}
   &  \multicolumn{4}{c}{$\hatRT{k}$}
   &  \multicolumn{4}{c}{$\hatS{k}$}
   &  \multicolumn{4}{c}{$\hatST{k}$}  \\[2ex]
\hline
$x$&$Q_{\lambda\mu} $&$=$&$ (-1)^{\lambda}    $&$Q_{\lambda,-\mu}
  $&$Q_{\lambda\mu} $&$=$&$ (-1)^{\lambda}    $&$Q_{\lambda,-\mu}
  $&$Q_{\lambda\mu} $&$=$&$                   $&$Q_{\lambda,-\mu}
  $&$Q_{\lambda\mu} $&$=$&$                   $&$Q_{\lambda,-\mu}    $\\
$x$&$M_{\lambda\mu} $&$=$&$ (-1)^{\lambda}    $&$M_{\lambda,-\mu}
  $&$M_{\lambda\mu} $&$=$&$-(-1)^{\lambda}    $&$M_{\lambda,-\mu}
  $&$M_{\lambda\mu} $&$=$&$-                  $&$M_{\lambda,-\mu}
  $&$M_{\lambda\mu} $&$=$&$                   $&$M_{\lambda,-\mu}    $\\[2ex]
$y$&$Q_{\lambda\mu} $&$=$&$ (-1)^{\lambda-\mu}$&$Q_{\lambda,-\mu}
  $&$Q_{\lambda\mu} $&$=$&$ (-1)^{\lambda-\mu}$&$Q_{\lambda,-\mu}
  $&$Q_{\lambda\mu} $&$=$&$ (-1)^{-\mu}       $&$Q_{\lambda,-\mu}
  $&$Q_{\lambda\mu} $&$=$&$ (-1)^{-\mu}       $&$Q_{\lambda,-\mu}    $\\
$y$&$M_{\lambda\mu} $&$=$&$ (-1)^{\lambda-\mu}$&$M_{\lambda,-\mu}
  $&$M_{\lambda\mu} $&$=$&$-(-1)^{\lambda-\mu}$&$M_{\lambda,-\mu}
  $&$M_{\lambda\mu} $&$=$&$-(-1)^{-\mu}       $&$M_{\lambda,-\mu}
  $&$M_{\lambda\mu} $&$=$&$ (-1)^{-\mu}       $&$M_{\lambda,-\mu}    $\\[2ex]
$z$&$Q_{\lambda\mu} $&$=$&$ (-1)^{\mu}        $&$Q_{\lambda  \mu}
  $&$Q_{\lambda\mu} $&$=$&$ (-1)^{\mu}        $&$Q_{\lambda  \mu}
  $&$Q_{\lambda\mu} $&$=$&$ (-1)^{\lambda+\mu}$&$Q_{\lambda  \mu}
  $&$Q_{\lambda\mu} $&$=$&$ (-1)^{\lambda+\mu}$&$Q_{\lambda  \mu}    $\\
$z$&$M_{\lambda\mu} $&$=$&$ (-1)^{\mu}        $&$M_{\lambda  \mu}
  $&$M_{\lambda\mu} $&$=$&$-(-1)^{\mu}        $&$M_{\lambda  \mu}
  $&$M_{\lambda\mu} $&$=$&$-(-1)^{\lambda+\mu}$&$M_{\lambda  \mu}
  $&$M_{\lambda\mu} $&$=$&$ (-1)^{\lambda+\mu}$&$M_{\lambda  \mu}    $\\[2ex]
\hline
$x$&$Q_{\lambda\mu} $&$=$&$ (-1)^{\lambda+\mu}$&$Q_{\lambda  \mu}^*
  $&$Q_{\lambda\mu} $&$=$&$ (-1)^{\lambda+\mu}$&$Q_{\lambda  \mu}^*
  $&$Q_{\lambda\mu} $&$=$&$ (-1)^{\mu}        $&$Q_{\lambda  \mu}^*
  $&$Q_{\lambda\mu} $&$=$&$ (-1)^{\mu}        $&$Q_{\lambda  \mu}^* $\\
$x$&$M_{\lambda\mu} $&$=$&$ (-1)^{\lambda+\mu}$&$M_{\lambda  \mu}^*
  $&$M_{\lambda\mu} $&$=$&$-(-1)^{\lambda+\mu}$&$M_{\lambda  \mu}^*
  $&$M_{\lambda\mu} $&$=$&$-(-1)^{\mu}        $&$M_{\lambda  \mu}^*
  $&$M_{\lambda\mu} $&$=$&$ (-1)^{\mu}        $&$M_{\lambda  \mu}^* $\\[2ex]
$y$&$Q_{\lambda\mu} $&$=$&$ (-1)^{\lambda}    $&$Q_{\ ambda  \mu}^*
  $&$Q_{\lambda\mu} $&$=$&$ (-1)^{\lambda}    $&$Q_{\lambda  \mu}^*
  $&$Q_{\lambda\mu} $&$=$&$                   $&$Q_{\lambda  \mu}^*
  $&$Q_{\lambda\mu} $&$=$&$                   $&$Q_{\lambda  \mu}^* $\\
$y$&$M_{\lambda\mu} $&$=$&$ (-1)^{\lambda}    $&$M_{\lambda  \mu}^*
  $&$M_{\lambda\mu} $&$=$&$-(-1)^{\lambda}    $&$M_{\lambda  \mu}^*
  $&$M_{\lambda\mu} $&$=$&$-                  $&$M_{\lambda  \mu}^*
  $&$M_{\lambda\mu} $&$=$&$                   $&$M_{\lambda  \mu}^* $\\[2ex]
$z$&$Q_{\lambda\mu} $&$=$&$ (-1)^{\mu}        $&$Q_{\lambda  \mu}
  $&$Q_{\lambda\mu} $&$=$&$ (-1)^{\mu}        $&$Q_{\lambda  \mu}
  $&$Q_{\lambda\mu} $&$=$&$ (-1)^{\lambda+\mu}$&$Q_{\lambda  \mu}
  $&$Q_{\lambda\mu} $&$=$&$ (-1)^{\lambda+\mu}$&$Q_{\lambda  \mu}   $\\
$z$&$M_{\lambda\mu} $&$=$&$ (-1)^{\mu}        $&$M_{\lambda  \mu}
  $&$M_{\lambda\mu} $&$=$&$-(-1)^{\mu}        $&$M_{\lambda  \mu}
  $&$M_{\lambda\mu} $&$=$&$-(-1)^{\lambda+\mu}$&$M_{\lambda  \mu}
  $&$M_{\lambda\mu} $&$=$&$ (-1)^{\lambda+\mu}$&$M_{\lambda  \mu}   $\\
\end{tabular}
\end{center}
\end{table}}]\narrowtext

{}For the conserved parity, one obtains the standard conditions:
   \bnll{eq703}
        Q_{\lambda\mu} &=& (-1)^\lambda Q_{\lambda\mu}
                                              , \label{eq703a} \\
        M_{\lambda\mu} &=&-(-1)^\lambda M_{\lambda\mu}
                                              , \label{eq703b}
   \enll%
i.e., odd electric and even magnetic moments must vanish.
Similar symmetry properties with respect to other symmetries
of the {\DT} or {\DTD} groups are collected in Table \ref{tab3}.

Within the standard phase convention of Eq.~(\ref{eq539}),
only a conservation of the $y$-$T$-simplex symmetry,
$\hatST{y}$, enforces
the reality of all multipole electric and magnetic moments.
In such a case, the lower part of Table \ref{tab3} gives at a glance
all the multipole moments which must vanish whenever any other
symmetry is additionally conserved; these are those
for which the phase factors are negative.
On the other hand,
a conservation of the $x$-$T$-simplex symmetry,
$\hatST{x}$, enforces the equality of
negative and positive magnetic components.
In this case, a conservation of any additional symmetry
puts to zero the multipole moments with
negative phase factors appearing in the upper part of the Table.
Of course, numerous other combinations of conserved symmetries
can be considered,
for example, a conservation of the $y$-simplex symmetry,
$\hatS{y}$, gives real electric moments and imaginary magnetic
moments.

Since conditions listed in Table \ref{tab3} depend only on the
parity of $\lambda$ and on the parity of $\mu$, and since
condition (\ref{eq539}) allows us to consider only non-negative
values of $\mu$, one has only six types of the symmetry properties of
multipole moments with respect to the {\DT} or {\DTD} operators.  These six
types are listed in Table \ref{tab5} for electric and magnetic
moments.  Column denoted by the identity operator $\hatE$ gives
the properties resulting solely from condition (\ref{eq539}),
while the remaining columns give properties of moments when one
of the non-identity {\DT} or {\DTD} operators is conserved.

In the same Table we also give symmetry properties of the
Cartesian components of the average angular momenta
$I_k$ (\ref{mvtam}).  Although the symmetry properties of
the angular momentum are identical to those of the dipole
magnetic moment, explicit values shown for its Cartesian
components allow for a simple visualization of a direction taken
by the angular-momentum vector when various {\DT} or {\DTD} operators are
conserved.  In particular, one can see that a conservation of
any of the signature or simplex operators for a given axis enforces the
angular-momentum direction along that axis, while a
conservation of any $T$-signature or $T$-simplex operators
allows for a tilted angular momentum in the plane perpendicular
to the given axis, cf.~Ref.\cite{[Fra00]}.
On the other hand, none of these operators
may be conserved if the angular momentum is to be tilted beyond
any of the $x$-$y$, $y$-$z$, and $z$-$x$ planes.  Note, however,
that the above tilting conditions pertain to the reference frame,
and not to the principal axes of the mass distribution.  An
appropriate choice of the reference frame, as discussed below,
has to be performed in
order to relate the conserved {\DT} or {\DTD} operators to the direction
of $I_k$ with respect to the mass principal
axes.

\twocolumn[
\mediumtext
{\begin{table}
{\caption[TT]{Properties of electric multipole moments
$Q_{\lambda\mu}$, magnetic multipole moments
$M_{\lambda\mu}$, and average angular momenta,
$I_k$ for conserved
{\DT} or {\DTD} operators. Symbols C, R, I, or 0 denote
values which can be, in general, complex, real, imaginary,
or zero, respectively.
\label{tab5}}}
\begin{center}
\begin{tabular}{c|cccccccccccccccc}
             & $\hatE   $  & $\hatT    $ & $\hatP   $  & $\hatPT   $ &
               $\hatR{x}$  & $\hatRT{x}$ & $\hatS{x}$  & $\hatST{x}$ &
               $\hatR{y}$  & $\hatRT{y}$ & $\hatS{y}$  & $\hatST{y}$ &
               $\hatR{z}$  & $\hatRT{z}$ & $\hatS{z}$  & $\hatST{z}$  \\
%
\hline
$Q_{10}$,$Q_{30}$,$Q_{50}$$\ldots$
         & R & R & 0 & 0 & 0 & 0 & R & R & 0 & 0 & R & R & R & R & 0 & 0 \\
$Q_{11}$,$Q_{31}$,$Q_{33}$$\ldots$
         & C & C & 0 & 0 & R & R & I & I & I & I & R & R & 0 & 0 & C & C \\
$Q_{20}$,$Q_{40}$,$Q_{60}$$\ldots$
         & R & R & R & R & R & R & R & R & R & R & R & R & R & R & R & R \\
$Q_{21}$,$Q_{41}$,$Q_{43}$$\ldots$
         & C & C & C & C & I & I & I & I & R & R & R & R & 0 & 0 & 0 & 0 \\
$Q_{22}$,$Q_{42}$,$Q_{44}$$\ldots$
         & C & C & C & C & R & R & R & R & R & R & R & R & C & C & C & C \\
$Q_{32}$,$Q_{52}$,$Q_{54}$$\ldots$
         & C & C & 0 & 0 & I & I & R & R & I & I & R & R & C & C & 0 & 0 \\
\hline
$M_{10}$,$M_{30}$,$M_{50}$$\ldots$
         & R & 0 & R & 0 & 0 & R & 0 & R & 0 & R & 0 & R & R & 0 & R & 0 \\
$M_{11}$,$M_{31}$,$M_{33}$$\ldots$
         & C & 0 & C & 0 & R & I & R & I & I & R & I & R & 0 & C & 0 & C \\
$M_{20}$,$M_{40}$,$M_{60}$$\ldots$
         & R & 0 & 0 & R & R & 0 & 0 & R & R & 0 & 0 & R & R & 0 & 0 & R \\
$M_{21}$,$M_{41}$,$M_{43}$$\ldots$
         & C & 0 & 0 & C & I & R & R & I & R & I & I & R & 0 & C & C & 0 \\
$M_{22}$,$M_{42}$,$M_{44}$$\ldots$
         & C & 0 & 0 & C & R & I & I & R & R & I & I & R & C & 0 & 0 & C \\
$M_{32}$,$M_{52}$,$M_{54}$$\ldots$
         & C & 0 & C & 0 & I & R & R & R & I & R & I & R & C & 0 & C & 0 \\
\hline
$I_x$
         & R & 0 & R & 0 & R & 0 & R & 0 & 0 & R & 0 & R & 0 & R & 0 & R \\
$I_y$
         & R & 0 & R & 0 & 0 & R & 0 & R & R & 0 & R & 0 & 0 & R & 0 & R \\
$I_z$
         & R & 0 & R & 0 & 0 & R & 0 & R & 0 & R & 0 & R & R & 0 & R & 0 \\
\end{tabular}
\end{center}
\end{table}}]\narrowtext

Independently of any {\DT} or {\DTD} symmetry breaking, the reference frame in
the space coordinates can be chosen in such way that some of the moments
have simple forms.  For example, a shift of the reference frame can
bring all electric dipole moments to zero (this corresponds to using
the center-of-mass system of reference), i.e.,
\be\label{eq303}
Q_{10}={\Re}Q_{11}={\Im}Q_{11}=0.
\ee
Similarly, a suitable
rotation of the reference frame can bring the electric quadrupole
moments $Q_{2\mu}$ to the principal axes, where \be\label{eq304}
{\Re}Q_{21}={\Im}Q_{21}={\Im}Q_{22}=0.  \ee

On the other hand, for some conserved symmetries, these
conditions can be automatically satisfied.  For example,
conservation of the {\Dh} group (i.e., simultaneous invariance
with respect to operators $\hatP$, $\hatR{x}$, and $\hatR{y}$)
ensures that the center-of-mass (\ref{eq303}) and
principal-axes (\ref{eq304}) conditions are automatically
satisfied, see Table \ref{tab5}.  Therefore, the breaking of the
{\DT} or {\DTD} symmetries may have non-trivial physical consequences
only for higher electric moments; starting from $Q_{30}$, if
the parity is broken, or starting from $Q_{41}$, for example,
if the parity is conserved.  In other words, the {\DT} or {\DTD}
symmetry breaking will not lead to new classes of
low-multipolarity shapes.  Nevertheless, such symmetry
breaking will immediately be reflected in values of magnetic
moments, whenever the time reversal is broken too.

\section{Conclusions}
\label{sec7}

In the present study we have presented applications of point groups
based on the three mutually perpendicular symmetry axes of the second
order, inversion, and time reversal, to nuclear structure problems.
We have discussed properties of the corresponding single {\DT}
and double {\DTD} groups in describing even and odd fermion systems,
respectively. We have enumerated their representations, both for
many-body states and for the single-particle operators, and reviewed
properties of group operators when they are represented in the fermion
Fock space.

Consequences of conserving individual {\DT} or {\DTD} symmetries have been
enumerated for:  (i) space symmetries of local one-body densities,
(ii) electric and magnetic multipole moments, and (iii) average
values of the angular-momentum operators. This gives information
about the nuclear shapes and matter-flow currents in states obeying
one or more of the {\DT} or {\DTD} symmetries, and allows for selecting
appropriate conserved symmetries in descriptions aiming at various
physical phenomena.

\acknowledgments

This research was supported in part by the Polish Committee for
Scientific Research (KBN) under Contract Nos.~2~P03B~034~08 and
2~P03B~040~14, and by the French-Polish integrated actions
programme POLONIUM.

\appendix
\section*{}
\label{appendix}

In this Appendix we explicitly construct irreducible representations
of the {\DTD} group by using the example of the HO basis
(Sec.~\ref{hoba}), and we illustrate the Wigner classification of
groups that contain antilinear operators (see Chap.~26 of
Ref.~\cite{[Wig59]}). The results of such an analysis were used in
Sec.~\ref{sec2}.

We consider here only the two-dimensional spinor representations,
appropriate for the odd-fermion systems and in particular for
the single-particle states. From Eqs.~(\ref{eq209})
one finds representation matrices $\Gamma(\hatMU)$ (where operators
$\hatMU$$\in${\DD} of Sec.~\ref{double} form
the double group {\DD}), in the two-dimensional invariant subspace spanned
by $\hoket{s_z\text{=}+\demi}$ and $\hoket{s_z\text{=}-\demi}$. We have
 \bnll{eqapp1}
  \Gamma(\,\hatME\,) &=& -\Gamma(\,\barME\,)
          =                                   \sigma_0  ,   \label{eqapp1a} \\
  \Gamma(\,\hatMP\,) &=& -\Gamma(\,\barMP\,)
          = (-1)^{n_x+n_y+n_z}                \sigma_0  ,   \label{eqapp1c} \\
  \Gamma(\hatMR{k})  &=& -\Gamma(\barMR{k})
          = -i(-1)^{N_k}                      \sigma_k  ,   \label{eqapp1e} \\
  \Gamma(\hatMS{k})  &=& -\Gamma(\barMS{k})
          = -i(-1)^{n_k}                      \sigma_k  ,   \label{eqapp1h}
 \enll
where $\sigma_0$ is the identity 2$\times$2 matrix, $\sigma_k$
for $k=x,y,z$ are the standard Pauli matrices,
and symbols $N_x$, $N_y$, and $N_z$ refer to $n_y$+$n_z$, $n_x$+$n_z$, and
$n_x$+$n_y$, respectively.

The characters of the classes are
 \bnll{eqapp2}
  \chi(\hatME)\,=\,-\chi(\barME)&=&2,                   \label{eqapp2a} \\
  \chi(\hatMP)=-\chi(\barMP)&=&2(-1)^{n_x+n_y+n_z}, \label{eqapp2b} \\
  \chi(\{\hatMR{x},\barMR{x}\})=\chi(\{\hatMS{x},\barMS{x}\}) &=& 0,
                                                    \label{eqapp2c} \\
  \chi(\{\hatMR{y},\barMR{y}\})=\chi(\{\hatMS{y},\barMS{y}\}) &=& 0,
                                                    \label{eqapp2d} \\
  \chi(\{\hatMR{z},\barMR{z}\})=\chi(\{\hatMS{z},\barMS{z}\}) &=& 0.
                                                    \label{eqapp2e}
 \enll
One can see, that only the characters of $\hatMP$ and $\barMP$ depend on
quantum numbers $n_x$, $n_y$, and $n_z$ that define the invariant
subspaces; more precisely, they depend only on the parity of the sum
$n_x+n_y+n_z$, i.e., on the total parity of basis states. Therefore, the
only two spinor representations of {\DTD}
can be labeled by the
eigenvalues of the parity operator $\hatMP$. Let us also note that all
characters are real.

If we introduce the time reversal, $\hatMT$, into the ensemble of
the linear operators belonging to {\DD} we obtain the {\DTD} group with 16
new antilinear elements $\hatMU^T$$\equiv$$\hatMU$$\hatMT$, Sec.~\ref{double}.
To study properties of the representations of
the {\DTD} group, one has to consider representations provided by
matrices
 \be\label{eqapp3}
    \breve\Gamma(\hatMU)=\Gamma(\hatMA^{-1}\hatMU\hatMA)^*,
 \ee
where $\hatMA$ is one of the antilinear elements of {\DTD} (see
\cite{[Wig59]}). It is most convenient to take $\hatMT$ itself as
$\hatMA$; we then have simply
 \be\label{eqapp4}
    \breve\Gamma(\hatMU)=\Gamma(\hatMU)^*,
 \ee
as $\hatMT$ commutes with all $\hatMU$$\in${\DD}. In such a case, matrices
$\breve\Gamma(\hatMU)$ are just complex conjugates of $\Gamma(\hatMU)$, and
therefore the characters of representation $\breve\Gamma$ are
exactly the same as those of $\Gamma$, because they are all real,
see Eqs.~(\ref{eqapp2}). Therefore these two representations are
equivalent, and a matrix $\beta$ exists which brings by a similarity
transformation all matrices $\Gamma(\hatMU)$ to $\breve\Gamma(\hatMU)$=$\Gamma(\hatMU)^*$,
  \begin{equation}
     \beta^{-1}\Gamma(\hatMU)\beta = \Gamma(\hatMU)^*, \quad \hatMU\in\text{\DD}.
  \end{equation}
Now, as shown by Wigner\cite{[Wig59]}, there are only two cases
possible: either
  \be\label{eqapp5}
    \beta\beta^*=+\Gamma(\hatMT^2),
  \ee
or
  \be\label{eqapp6}
    \beta\beta^*=-\Gamma(\hatMT^2).
  \ee
Matrix $\beta$ can easily be found from the explicit expressions
for matrices $\Gamma(\hatMU)$ given in Eqs.~(\ref{eqapp1}), and it reads
  \be\label{eqapp7}
      \beta=e^{i\phi}\left(\ba{rr}  0 & -1 \\
                                    1 &  0 \ea
           \right)  = -ie^{i\phi}\sigma_y .
  \ee
Choosing the phase factor $e^{i\phi}$$\neq$1 in (\ref{eqapp7})
is equivalent to a change of phase of the $\hoket{s_z\text{=}+\demi}$
states, and to a change in the phase convention in
Eq.~(\ref{eq209b}).

It is easy to demonstrate that with this form of the matrix $\beta$,
Eq.~(\ref{eqapp5}), and not (\ref{eqapp6}) holds.
In Wigner's classification this case leads to what is called the
corepresentations of the
``first kind'': any representation $\Gamma$ of the group {\DD} can be
completed to a corepresentation of the full {\DTD} group by defining
  \be\label{eqapp8}
     \Gamma(\hatMU^T)=\Gamma(\hatMU)\beta.
  \ee
Note that taking $\hatMU$=$\hatME$ one gets $\Gamma(\hatMT)$=$\beta$,
so $\beta$ is, of course, just the matrix representing $\hatMT$
itself.

After Wigner, the term {\em corepresentation} is used here because the
representations of groups containing antilinear operators are
{\em not} representations in the usual sense. To see this, let us consider
an orthonormal set of states $\lbrace|\phi_i\rangle\rbrace$ constituting
a basis of a representation $\Gamma$. Let $\hatMU$ be any linear, and $\hatMU'$
any element of the group. Then, because of {antilinearity} of $\hatMU^T$
one has
  \bn\label{eqapp9}
    (\hatMU^T\cdot\!\hatMU')|\phi_j\rangle&=&\sum_i \hatMU^T \Gamma(\hatMU')_{ij}|\phi_i\rangle
           =\sum_i \Gamma(\hatMU')_{ij}^* \hatMU^T|\phi_i\rangle    \nonumber    \\
    &=& \sum_{ik} \Gamma(\hatMU')_{ij}^*\Gamma(\hatMU^T)_{ki}|\phi_k\rangle
                                                         \nonumber    \\
    &=& \sum_{k} \left[\Gamma(\hatMU^T)\Gamma(\hatMU')^*\right]_{kj}
                                           |\phi_k\rangle
  \en
and, consequently,
  \be\label{eqapp10}
    \Gamma(\hatMU^T\cdot\!\hatMU') = \Gamma(\hatMU^T)\Gamma(\hatMU')^*,
  \ee
to be compared with
  \be\label{eqapp11}
    \Gamma(\hatMU\cdot\!\hatMU') = \Gamma(\hatMU)\Gamma(\hatMU'),
  \ee
which holds for the ``usual'' representations.
The presence of complex conjugation on the right-hand
side of Eq.~(\ref{eqapp10}) implies that the homomorphism
between the group
multiplication and the multiplication of representation matrices no longer
holds when the group contains antilinear operators.
This is not surprising in view of the fact that matrices,
by construction, always act on vectors (columns of numbers) linearly.

In conclusion, there are only two spinor corepresentations of {\DTD}, and
they can be labeled, as is also the case for the {\DD} group, by one quantum
number only (parity).


\begin{thebibliography}{10}

\bibitem{[RS80]}
{P. Ring and P. Schuck, {\sl The Nuclear Many-Body Problem} (Springer-Verlag,
  Berlin, 1980)}.

\bibitem{[Ham62]}
{M. Hamermesh, {\em Group Theory} (Addison-Wesley, Reading, Mass., 1962)}.

\bibitem{[Kos63]}
{G.F. Koster, J.O. Dimmock, R.G. Wheeler and H. Statz,
{\sl Properties of the Thirty-Two Point Groups} (M.I.T. Press, Cambridge,
Massachusetts, 1963)}.

\bibitem{[Lan81]}
{L.D. Landau and E.M. Lifschitz, {\sl Quantum Mechanics (Non-relativistic
Theory)} (Pergamon Press, 1981)}.

\bibitem{[Cor84]}
{J.F. Cornwell, {\sl Group Theory in Physics}  (Academic Press, 1984)}.

\bibitem{[Cur62]}
{Ch.W. Curtis, I. Rainer, {\sl Representation Theory of Finite Groups and
Associative Algebras} (Interscience Publishers, New York, 1962)}.

\bibitem{[Bra72]}
{C.J. Bradley, A.P. Cracknell,
{\sl The Mathematical Theory of Symmetry in Solids} (Clarendon Press, Oxford,
1972)}.

\bibitem{[Wig59]}
{E.P. Wigner,  {\sl Group Theory and Its Application to the Quantum
 Mechanics of Atomic Spectra} (Academic Press, New York, 1959)}.


\bibitem{[Bon87]}
{P. Bonche, H. Flocard, and P.-H. Heenen, Nucl. Phys. {\bf A467}, 115-35
  (1990)}.

\bibitem{[Dob97]}
{J. Dobaczewski and J. Dudek, Comp. Phys. Commun. {\bf 102}, 166 (1997); {\bf
  102}, 183 (1997)}.

\bibitem{[Neg75]}
{J.W. Negele, {\em Lecture Notes in Physics 40} (Springer, Berlin, 1975) pp.
  285, 288}.

\bibitem{[Dre90]}
{R.M. Dreizler and E.K.U. Gross, {\em Density Functional Theory} (Springer,
  Berlin, 1990)}.

\bibitem{[Que78]}
{P. Quentin and H. Flocard, Annu. Rev. Nucl. Part. Sci. {\bf 28}, 523 (1978)}.

\bibitem{[Dob00b]}
{J. Dobaczewski, J. Dudek, S.G. Rohozi\'nski, and T.R. Werner, the following
paper}.

\bibitem{[Mes62]}
{A. Messiah, {\em Quantum Mechanics} (Wiley, New York, 1962)}.

\bibitem{transposition}
{In this article, superscript $T$
always means that the corresponding operator contains factor $\hatT$;
this should not be confused with the notation for the transposition
operation.}

\bibitem{[Gir83]}
{M. Girod and B. Grammaticos, Phys. Rev. {\bf C27}, 2317 (1983)}.

\bibitem{Var-Bohr}
{The phase convention implied by Eq.~(\protect\ref{eq209b}) agrees
with that of Varshalowitch {\it et al.} \protect\cite{[Var88]}, and thus it is
opposite to the one used by Bohr and Mottelson \protect\cite{[Boh69]}.}

\bibitem{[Rip68]}
{G. Ripka, Adv. Nucl. Phys. {\bf 1}, 183 (1968)}.

\bibitem{[Eng75]}
{Y.M. Engel, D.M. Brink, K. Goeke, S.J. Krieger, and D. Vautherin, Nucl. Phys.
  {\bf A249}, 215 (1975)}.

\bibitem{[Var88]}
{D.A Varshalovitch, A.N. Moskalev, and V.K. Kersonskii, {\sl Quantum Theory of
  Angular Momentum} (World Scientific, Singapore, 1988)}.

\bibitem{phacon}
{We adopt the general phase convention \protect\cite{[Var88]} valid for
  arbitrary irreducible tensor operators. However, note that for
  the integer-angular-momentum irreps, the signs of magnetic
  components appearing in the phase
  factors can be arbitrarily changed.}

\bibitem{[Fra00]}
{S. Frauendorf, to appear in Reviews of Modern Physics}.

\bibitem{[Boh69]}
{A. Bohr and B.R. Mottelson {\sl Nuclear Structure (Vol. I)}
(W.A. Benjamin, Inc., 1969), p.19}.

\end{thebibliography}

\end{document}